\documentclass[preprint,pra,superscriptaddress,floatfix,showpacs]{revtex4-2}
\usepackage[utf8]{inputenc}
\usepackage{amsmath}
\usepackage{amssymb}
\usepackage{graphicx}

\makeatletter
\usepackage{graphics}
\usepackage{epsfig}
\usepackage{color}

\newcommand{\be}{\begin{equation}} \newcommand{\ee}{\end{equation}}
\newcommand{\bea}{\begin{eqnarray}} \newcommand{\eea}{\end{eqnarray}}

\makeatother

\begin{document}

\title{Ultra-slow dynamics of free-running ring lasers: \\ towards a minimal model}

\author{Giovanni Giacomelli}
\affiliation{Consiglio Nazionale delle Ricerche, Istituto dei Sistemi Complessi, via Madonna del Piano 10, I-50019 Sesto Fiorentino (FI), Italy}
\author{Antonio Politi}
\affiliation{Consiglio Nazionale delle Ricerche, Istituto dei Sistemi Complessi, via Madonna del Piano 10, I-50019 Sesto Fiorentino (FI), Italy}
\affiliation{Department of Physics, Old Aberdeen, Aberdeen AB24 3UE, United Kingdom}

\date{\today}

\begin{abstract}
The dynamics of a resonant, free-running ring laser, in the common case of a fast relaxation of the atomic polarization, 
is unexpectedly highly singular. 
As shown in [Phys. Rev. Research, {\bf 5}, 023059 (2023)], this is due to the closeness to a pure Hamiltonian dynamics ruled by a nonlinear wave equation, herein named Klein-Gordon-Toda model. 
In this paper, we derive a quasi-Hamiltonian model which allows describing realistic systems. 
In particular, we identify two nearly conserved, energy-like quantities, which ``naturally" 
exhibit an ultra-slow dynamics confirmed and highlighted by numerical simulations. 
A minimal version of the quasi-Hamiltonian model is finally derived, which does not only reproduce the laser thresholds, but also helps understanding the origin of the nearly integrable character of the laser dynamics.
\end{abstract}

\maketitle

\section{Introduction}

A correct description of laser dynamics requires accounting for the evolution of the population inversion as well as of
the atomic polarization (besides obviously describing field propagation).
It is almost 60 years that an appropriate model has been derived for two-level atoms: these are the Arecchi-Bonifacio equations ~\cite{Arecchi1965}
(often cited as Maxwell-Bloch). 
They have been originally derived for the description of a coherent optical  amplifier, but they are also 
commonly used, with a proper choice of the boundary conditions,
in the study of free-running (a coherent amplifying medium only inside the cavity) resonant and unidirectional ring lasers. 
In the following, we will refer to this approach as the AB (ring laser) model.   
In such a system the first threshold, leading to a non-zero stationary field, has been since long determined, as well as the Risken-Nummedal-Graham-Haken (RNGH) threshold above which, more or less irregular oscillations set in~\cite{Risken1968,Risken1968a,Graham1968}. Sometimes the latter is called second threshold, even if this denomination is traditionally reserved to the onset  
of instabilities in low-dimensional contexts, where a single or a few modes are active
(see, e.g.,~\cite{Milonni1987} for a review of low-dimensional chaos). 
In the context of extensive (high-dimensional) dynamics, a normal form has been derived 
in the vicinity of the RNGH threshold and numerically validated~\cite{Casini1997}.

The existence of a mathematical model may not suffice to draw reliable conclusions on the behavior of a free-running ring laser.
In fact, the presence of multiple scales may hinder direct simulations, by requiring too small integration
time-steps and/or too-long integration times.
Whenever the presence of short relaxation times implies a fast convergence to some center manifold, the computational
complexity can be reduced by adiabatically eliminating one or more variables thereby allowing for larger integration
time steps.

Lasers characterized by a fast relaxation of the atomic polarization belong to a particularly nasty 
category of dynamical systems. This includes the broad classes of semiconductor lasers (see e.g. \cite{agrawal2013}) and doped-fiber based  system~\cite{Turitsyna2013,Rogers2005,Abarbanel1999,Ray2008}.
On the one hand, direct simulations of the original model are unfeasible.
On the other hand, the atomic polarization cannot be straightforwardly eliminated, as the resulting model is improperly unstable.
An alternative solution has been proposed via e.g. the introduction of the Haus master equation~\cite{Haus1975}, 
to describe mode-locking phenomena associated to pulse-propagation in a ring-cavity laser. 
However, although this approach proves effective 
in the context it was designed for, when used to describe the laser alone is unable even
to reproduce the ``second" laser threshold, thus missing a crucial feature of the  dynamics~\cite{Perego2020}.
Another model, the so-called Vladimirov-Turaev equation, specifically developed for semiconductors, 
is again unable to reproduce the RNGH laser threshold~\cite{Vladimirov2005} in the free-running laser setup.

The only model we are aware of, able to display the RNGH threshold is the so-called coherent master equation
(CME) derived in Ref.~\cite{Perego2020}, after performing an improved (second-order) adiabatic elimination.
This model leaves, however, open the question of why it is so difficult even to reproduce qualitatively
the correct behavior.

A first step towards the solution of this puzzle was made in~\cite{Giacomelli2021}, where the spatial dependence of the
 AB model was eliminated, transforming it into a purely delayed equation
(the success of the approach being due to the synchronization of the field in a moving frame with the input field
- see \cite{Giacomelli2021} for a more detailed explanation).

Although this simplification was not sufficient to enable realistic simulations of e.g. fiber lasers, 
it opened the venue towards substantial progress, thanks to the introduction of a powerful perturbative technique
based on the smallness parameter $\Gamma = \sqrt{\gamma_\parallel/\gamma_\perp}$, where 
$\gamma_\parallel$ and $\gamma_\perp$ are the decay rates of the atomic population and polarization,
respectively.
 In fact, in Ref.~\cite{Politi2023}
it was shown that the adiabatic elimination of a newly introduced variable
(a suitable combination of atomic polarization and electric field) leads, to first order,
to a Hamiltonian dynamics. We name it
Klein-Gordon-Toda (KGT), since it reduces to the Klein-Gordon model in the linear limit, and is characterised by a Toda-like \cite{Toda1975} 
exponential nonlinearity.  The KGT model possesses two conservation laws, called potential and kinetic energies
(for obvious reasons, once their definition will be given).

In this paper, we complete the project. By determining the next-order correction terms, we obtain
a model where the energies are self-determined and turn out to evolve over extremely long time scales. We call it
quasi-Hamiltonian model (QHM).
The QHM is very similar to the above mentioned CME.
However, the absence of a proper perturbative background behind the derivation of the CME prevents recognizing: (i) the closeness
to a Hamiltonian dynamics, (ii) that some terms are negligible and finally (iii) the presence of ultra-slow quasi-conserved quantities.

The QHM allows for realistic simulations of fiber lasers, especially thanks to the identification of an appropriate slow
time scale. The effectiveness of the model is related to the unexpectedly small fluctuations exhibited by the field intensity
even much above threshold. We provide an explanation, exploiting the interpretation of the laser regime
as a nonequilibrium steady state.
Starting from the '70s of the past century, the laser was recognized to be a valuable instance 
of stationary nonequilibrium state, since it is a device where an incoming energy flux
(necessary to ensure a steady population inversion) is partially dissipated via both atomic collisions and
imperfect mirrors, but also transformed into coherent light (the laser emission).
In the present context, the interpretation is entirely different: the
energy formally corresponds to the Hamiltonian of the KGT system (unrelated to the true physical
energy), while the incoming and outgoing fluxes are associated to suitable perturbative terms 
which destabilize/stabilize the various  modes.

The AB model includes explicitly three different scales: (i) the relaxation time 
$1/\gamma_\perp$ of the atomic polarization; (ii) the decay time $1/\gamma_\parallel$ of the population
inversion; the round-trip time $\mathcal{T}$ of the cavity.
The QHM shows that the relevant times scales exhibited during the dynamical evolution are
(besides $\mathcal{T}$): 
(1) the period $1/\sqrt{\gamma_\perp \gamma_\parallel}$ of the self-generated fast oscillations above the RNGH threshold;
(2) the Hamiltonian time scale $\mathcal{T}\sqrt{\gamma_\perp/\gamma_\parallel}$, which describes the evolution of
the coherent pseudo-spatial structures found in a delay interval (see \citep{Politi2023}); 
(3) the time scale $\mathcal{T}\gamma_\perp/\gamma_\parallel$ of energy variations.
This, without including the subtle scale associated
to the ergodization time of the KGT dynamics, very long, as the laser turns out to operate in a nearly integrable
regime.

The paper is organised as follows. In Sec.~\ref{sec:zero}, starting from the delayed version of the AB model,
we implement the first step of a perturbative approach which, via the introduction of a space-time representation, 
leads to the KGT model (the relationship between the AB model and its delayed version is briefly 
recalled in appendix \ref{app0}).
This is essentially a reformulation of~\cite{Politi2023}; it is a necessary step to set the formalism and to define
the proper background for the next relevant stage.

Sec.~\ref{sec:qhm} contains the core of the model derivation (the most technical calculations
are presented in appendix~\ref{app1}). The full QHM model contains several terms which are difficult
to analyse and interpret. Therefore, we have decided to focus on a simpler version, still accurate
for high reflectivity and relatively long round-trip time, conditions satisfied
in typical experimental setups.
In the same section, we perform the linear stability analysis, which reveals the smallness
of all eigenvalues, and present the first simulations to illustrate the properties of the expected
dynamics. In particular, we investigate the dependence of the stationary state on the pump value.

In Sec.~\ref{sec:cme}, we compare the QHM model with the CME, rewritten in our notations 
to identify the order of magnitude of the various terms.
The most important difference is perhaps the presence, in the CME, of an evolution
equation for the (pseudo-spatial) average of the population inversion, absent in our model.
We justify this result, by showing that such a variable can be adiabatically eliminated, as implicitly done
within our formalism. Finally, we show that some terms of the CME are negligible. 

In Sec.~\ref{sec:energy}, we discuss the energy dynamics
for different values of the smallness parameter, providing an argument to explain
the observed decrease of the average energy with $\Gamma$.
Therein, we also provide a theoretical justification for the self-selection of the ratio between
kinetic and potential energy.
The last section is devoted to a brief summary of the main results, to 
a presentation of the open problems, and to a discussion of future perspectives.

\section{Klein-Gordon-Toda model} \label{sec:zero}

As briefly summarized in the appendix \ref{app0}, the AB model is well approximated by the
delayed dynamical system
\begin{eqnarray}
F &=& F^d+ \Gamma \frac{1-R}{R}U \label{orig1} \\
 \dot U &=& \frac{R}{\Gamma} [-U + GF - \dot F^d] \label{orig2} \\
\dot G &=& -\Gamma G + I(1- F^2 -\Gamma FU) \label{orig3}  \; .
\end{eqnarray}
where $F$ is the field amplitude ($F^d \equiv F(t-T)$, $T$ being the roundtrip time in the actual time units), 
$G$ the population inversion, and $U$ is an auxiliary variable that can be
adiabatically eliminated. 
Moreover, $R$ is the mirror reflectivity, while $I$ is the rescaled pump parameter, herein denoted as ``effective" pump.
From its definition (\ref{eq:pump}), given in the appendix \ref{app0}, 
it turns out that the first laser threshold is $I=0$, which corresponds to a pump value $a = 1-R$
in the original formulation.

From the point of view of nonlinear dynamics,
this model is peculiar, since the delayed variable contributes both as itself and via its time derivative.
This means that the equation is neutrally stable (see e.g. \cite{bainov1991oscillation}), 
i.e. it belongs to a not-yet
fully understood class of dynamical systems.

The most relevant laser systems are characterized by a very small $\Gamma$.
It is, therefore, tempting to implement a perturbative approach, by expanding in powers of $\Gamma$.
In the following, the order of approximation
of the three variables $F$, $G$, and $U$ is identified by a
subscript number.

For $\Gamma \to 0$, we assume that all variables stay finite:
this hypothesis has been verified by running several simulations
for different $\Gamma$ values (see~\cite{Politi2023}), and it
is confirmed a posteriori by the consistency of the results. 
Hence, at zero order, we are authorized to set $\Gamma=0$ 
in Eqs.~(\ref{orig1},\ref{orig3}), finding that they reduce to
\begin{eqnarray}
F_0 &=& F_0^d \label{cond1} \\
\dot G_0 &=& I(1-F_0^2) \label{cond2}
\end{eqnarray}
while the variable $U$ turns out to be irrelevant. 

Condition (\ref{cond1}) implies that any
periodic function $F_0(t)$ of period $T$ is a valid solution, provided that
\[
\int_0^T dt F_0^2(t) = 1
\]
which follows from Eq.~(\ref{cond2}), since also $G_0$ must be periodic. 
This result shows that the original model is close to a very degenerate scenario.

At first order in $\Gamma$, all terms in Eqs.~(\ref{orig1},\ref{orig3}) must be retained
\begin{eqnarray}
F_1 &=& F_1^d + \frac{1-R}{R}\Gamma U_1 \label{eq:new1} \\
\dot G_1 &=& -\Gamma G_1 + I(1- F_1^2 -\Gamma FU_1)  \label{eq:new2}
\end{eqnarray}
showing that the $U$ dynamics must be now included.
Since $U$ is characterized by a fast relaxation, we can perform a standard adiabatic
elimination, setting its time derivative equal to zero 
\begin{equation}
U_1 = G_1F_1 -\dot F_1^d  \label{ae1}
\end{equation}
Upon inserting this $U_1$ value into Eq.~(\ref{eq:new1}), we obtain 
\begin{equation}
F_1 = F_1^d + \varepsilon ( G_1F_1 -\dot F_1^d ) \label{eq:neww1}
\end{equation}
where we have defined the new smallness parameter,
\begin{equation}
  \varepsilon = \frac{1-R}{R} \Gamma \; .
\end{equation}
We will generically refer to both parameters 
as the ``smallness'' parameter, selecting the definition we deem more appropriate in each given context.

Eqs.~(\ref{eq:new2},\ref{eq:neww1}) provide a first self-consistent representation of the laser model.
One should insert the $U_1$ expression also into Eq.~(\ref{eq:new2}), but we leave it implicit, as we are
going to show that such term is not relevant at this order of approximation.

\begin{figure}
\includegraphics[width=0.5\textwidth]{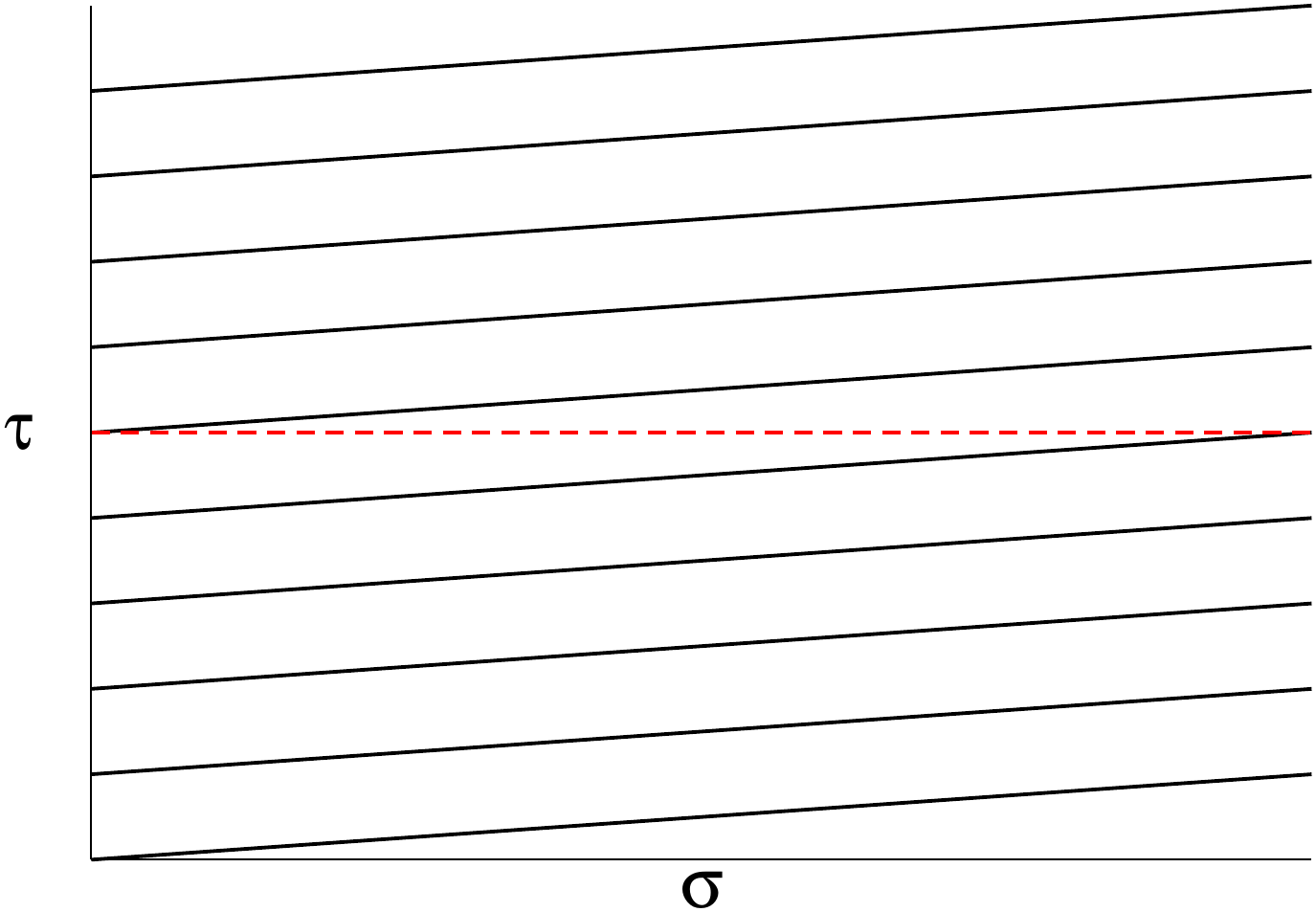}
   \caption{Space-time representation: the solid black line represents the initial one-dimensional time axis, 
which we imagine wrapped around a circular cylinder.  
$\sigma$ and $\tau$ increase from left to right and from bottom to top respectively. The dashed line corresponds to a constant $\tau$ line,
obtained via an interpolation from two neighboring world lines.}
   \label{fig:a}
\end{figure}

The model dynamics is better understood
by introducing the space time representation $t \to (\sigma,\tau)$ shown in Fig.~\ref{fig:a}, 
\begin{eqnarray}
\sigma = t \mod{T} \label{eq:sigma}\\
\tau = \varepsilon (\textrm{Int}(t/T) + \sigma/T) \label{eq:tau}
\end{eqnarray}
The space-like variable $\sigma$ identifies the position along a single delay unit; $\tau$
is a new time variable which quantifies the rescaled number of elapsed delay units.
This is a refined version of the spatio-temporal representation introduced in Ref.\cite{Arecchi1992} and first used in
 Ref.~\cite{Giacomelli1996}
to derive the normal form of delayed equation. We remark that such representation is valid in general, and not limited to the case of long delays as in the above references.

For the sake of simplicity, the two fields $F$ and $G$ are denoted in the same way as before, as
the number of arguments suffices to identify the underlying system of reference.
The spatial boundary conditions are periodic (here we drop the subscript, since the conditions is
valid at any approximation order):
$F(\sigma=0,\tau) = F(\sigma=T,\tau)$ (and analogously for $G$) 

In view of the small variation of $F$ from one to the next delay unit, we assume the $\tau$ dependence to be 
effectively continuous (and smooth), and write
\begin{equation}
F^d = F(\sigma, \tau-\varepsilon)\approx F -\varepsilon \partial_\tau F + 
\frac{\varepsilon^2}{2} \partial_\tau^2 F \; . \label{st1}
\end{equation}
Moreover, from Eq.~(\ref{eq:tau}), 
the original time derivative of a generic variable $H(t)$ can be expressed in terms of the new variables as
\begin{equation}
\dot H = \partial_\sigma H + \frac{\varepsilon}{T} \partial_\tau H \label{st2}
\end{equation}

With the help of Eqs.~(\ref{st1},\ref{st2}) (and retaining only terms linear in $\varepsilon$),
Eq.~(\ref{eq:neww1}) reduces to
\begin{equation}
\partial_\tau F_1 + \partial_\sigma F_1= G_1F_1 \label{ord1a}
\end{equation}
Notice that the explicit $\varepsilon$ dependence disappears, since all leading terms are of order $\varepsilon$.

By proceeding in a similar way with Eq.~(\ref{eq:new2}) and retaining the leading terms
(which, in this case, are of $\mathcal{O}(1)$), the second equation can be written as (notice that it
is equivalent to the zero-order approximation),
\begin{equation}
\partial_\sigma G_1 = I(1- F_1^2) \; .  \label{ord1b}
\end{equation}

By dropping the subscripts and introducing lower-case notations, 
the model equations are written as
\begin{eqnarray}
	\partial_\tau f &=& -\partial_\sigma f + g f  \label{approx1}  \\
	\partial_\sigma g &=& I\left(1- f^2\right) \; .  \label{approx2} 
\end{eqnarray}
By introducing $s = \ln f$, the model can be rewritten as
(in a moving frame with unit velocity)
\begin{equation}
\partial_{\tau\sigma}s = I(1-{\rm e}^{2s})  \; ,
\label{eq:hams}
\end{equation}
where, to avoid unnecessary complications, the spatial variable is still denoted with $\sigma$.
It can be easily verified that these equations are parameterless, since the effective pump $I$ can be eliminated 
by suitably rescaling the system scales: time ($\tau' = \tau \sqrt{I}$)), space ($\sigma' = \sigma \sqrt{I}$), and the population inversion ($g' = g/\sqrt{I}$). Thus, the system-energy gain (pump) and the losses (dissipation) only contribute to the extensivity of the dynamics.

In Ref.~\cite{Politi2023}, it was found that Eqs.~(\ref{approx1},\ref{approx2}) describe a perfectly Hamiltonian dynamics, 
characterized by two conservation laws.

The first invariant is the kinetic-like energy density
\begin{equation}
E_K =  \frac{\langle s^2_{\sigma'}\rangle}{2}  = \frac{\langle s^2_\sigma \rangle}{2I}  \, ;
\label{eq:ekin}
\end{equation}
here and in the following, the angular brackets denote a spatial average.
Since in this paper we will keep referring to $\tau$ and $\sigma$, we will always refer to the rightmost definition 
of kinetic energy.

The second conserved quantity is the pseudo-potential energy density
\begin{equation}
E_P = \langle {\rm e}^{2s} -2s -1 \rangle  \equiv  \langle V(s) \rangle  \; ,
\label{eq:epon}
\end{equation}
where $V(s)$ is a Toda-like potential, which provides a nonlinear contribution to the overall dynamics. 
{\it Cela va sans dire}, these energies have nothing to do with the physical energy pumped into the laser and partially 
transformed into electromagnetic fields. 

Altogether, the total energy (density), is
\begin{equation}
E_T = E_K + E_P
\end{equation}
evidently conserved. Its value quantifies the strength of the nonlinearity. For $E_T \ll 1$, the dynamics is 
effectively linear and thereby integrable (i.e. there are infinitely many conserved quantities).
In this limit, the model reduces to the Klein-Gordon equation. Since nonlinearities arise from the Toda-like
potential, we name Eqs.~(\ref{approx1},\ref{approx2}) Klein-Gordon-Toda (KGT) model.

For relatively large $E_T$ values, one expects chaos to set in, accompanied by more or less fast relaxation phenomena.
The value of the two energies $E_K$ and $E_P$ is fixed by the initial condition; hence, even if the model is parameterless,
it is necessary to fix these two quantities to determine the resulting dynamics.
In the next section we will see that higher-order terms break the Hamiltonian structure and contribute to select the values of
the ``conserved" quantitities, otherwise undetermined.

Numerically, given the current amplitude profile $f(\sigma,\tau)$, one starts integrating Eq.~(\ref{approx2}) (in space) 
to generate 
\begin{equation}
g(\sigma,\tau) = I [\sigma -  \widehat{f^2} (\sigma,\tau)] + C_1(\tau)
\label{eq:g}
\end{equation}
where the hat denotes a generic integral over $\sigma$ of the underlying function, while $C_1$ is a (yet unknown)
integration constant.

Periodic boundary conditions require that
\begin{equation}
\widehat{f^2}(T,\tau)= T  \qquad \Leftrightarrow \qquad \langle f^2 \rangle = 1 \; .
\end{equation}
Since this condition must be valid at all times,
the time derivative of $\langle f^2\rangle$ must be constant.
From the dynamical equations (\ref{approx1},\ref{approx2}),
\begin{eqnarray}
0 = \partial_\tau \langle f^2 \rangle &=& 
\frac{2}{T} \int_0^T gf^2 d\sigma = \frac{2}{T} \int_0^T \left( g -\frac{1}{I}g \partial_\sigma g\right)d\sigma
\nonumber \\
&=&  \frac{2}{T}\int_0^T g d\sigma= \tilde{g}(0,\tau) ~, 
\end{eqnarray}
where $\tilde{g}(0,\tau)$ is the zeroth-order Fourier mode. 
Therefore, selecting $C_1(\tau)$ such that $\tilde g(0,\tau) = 0$ implies that 
$\langle f^2 \rangle_\tau =0$ at all times.

The model can be integrated by expanding the field in Fourier modes.
By using a sufficiently large number of modes (typically, $N=2^{14}$ suffices) no numerical instability
emerges and there is no need to include an empirical smoothing as previously done in~\cite{Politi2023}.

\section{Quasi-Hamiltonian model} \label{sec:qhm}

The KGT model is characterized by two conserved quantities that we have called potential and kinetic energy: 
their value is encoded in the initial conditions.  Afterwards, they are left invariant by the dynamics.
Higher-order perturbative terms contribute to the selection of the
two energy values as well as to determine their stability properties.

A general second-order model accounting for all relevant corrections is derived in appendix~\ref{app1}.
Given its computational complexity, here we consider the simplified version, which emerges in the limit
$1-R\ll 1$ and for not too-short delay $T\gg 1-R$. Both approximations are valid in realistic physical conditions.
Under the above assumptions, Eqs.~(\ref{eq:finalf}, \ref{eq:finalg}) reduce to
\begin{equation}
\partial_\tau f +\partial_\sigma f = gf - \Gamma [ If(1-f^2)  + g\partial_\sigma f  - \partial_\sigma^2 f]
\label{eq:finalfs}
\end{equation}

\begin{equation}
\partial_\sigma g = I(1-f^2) - \Gamma \left [ g + I f(gf - \partial_\sigma f ) \right ]
\label{eq:finalgs}
\end{equation}
where, for the sake of simplicity,  we have reintroduced the smallness parameter $\Gamma$.

The term $\partial_\sigma f$ in Eq.~(\ref{eq:finalgs}) can be eliminated by choosing a frame moving with velocity 1.
To avoid another change of notations, we keep using $\sigma$ to denote the spatial position.
A last simplification can be made by noticing that the term $I(1-f^2)$ multiplying $f$ in Eq.~(\ref{eq:finalfs}),
can be replaced by $\partial_\sigma g$ (see Eq.~(\ref{eq:finalgs})) without lowering the order of
approximation. Hence, the field equation can be written as

\begin{equation}
\partial_\tau f = gf - \Gamma [ \partial_\sigma (fg)  - \partial_\sigma^2 f]
\label{eq:finalfs2}
\end{equation}

Eqs.~(\ref{eq:finalgs}, \ref{eq:finalfs2}) represent the final version of the  quasi-Hamiltonian model (QHM) 
we are going to investigate in the following.

\subsection{Linear stability} \label{sec:linestab}
Here, we show that the QHM reproduces the stability of
the AB model. We proceed by assuming
\begin{equation*}
f = 1+ q 
\end{equation*}
so that $q = g =0$ corresponds to the stationary lasing state.
By linearizing Eqs.~(\ref{eq:finalgs},\ref{eq:finalfs2}), we obtain
\begin{equation}
\partial_\tau q = g + 2\Gamma \partial_\sigma g  + \Gamma \partial_\sigma^2 q
\label{eq:qlin}
\end{equation}

\begin{equation}
\partial_\sigma g = -2Iq -\Gamma(1+I) g + \Gamma I \partial_\sigma f
\label{eq:glin}
\end{equation}
Now, introducing the Ansatz
\begin{equation}
  q =  q_0 \exp(\lambda t + i k\sigma) \qquad,\qquad~
  g =  g_0 \exp(\lambda t + i k\sigma)~,
\end{equation}
the above equations can be rewritten as

\begin{equation}
\lambda q_0 = g_0 - \Gamma (ikg_0+k^2q_0)  
\end{equation}

\begin{equation}
ik g_0 = -2Iq_0 -\Gamma(1+I) g_0 + ik \Gamma I q_0
\end{equation}
From this equation
\begin{equation}
g_0 = I \frac{-2+ik\Gamma}{ik +\Gamma(1+I)} q_0
\end{equation}
Hence, neglecting $\Gamma^2$ terms
\begin{equation}
\lambda \simeq I \frac{-2+3ik\Gamma}{\Gamma(1+I)+ik} -\Gamma k^2 \simeq 
 I \frac{-2(1+I)\Gamma +3 \Gamma k^2 + 2ikI}{k^2} -\Gamma k^2  
\end{equation}
so that the real part of $\lambda$ is

\begin{equation}
\frac{\lambda_R}{\Gamma} \simeq -\frac{2I(1+I)}{k^2} + 3I  -k^2
\end{equation}
The rates are all of order $\Gamma$ (independently whether stable or unstable).
\begin{figure}
\includegraphics[width=0.5\textwidth]{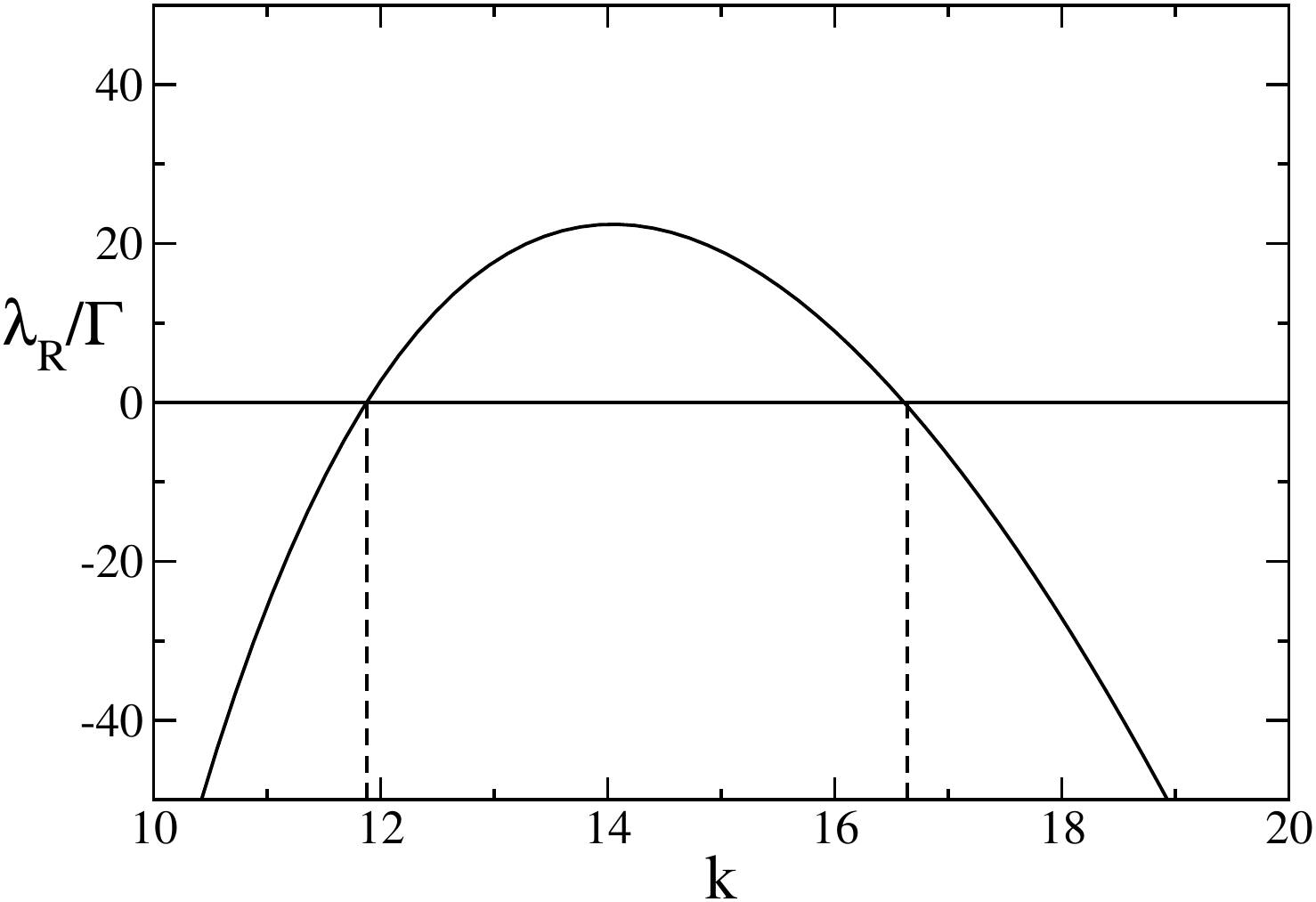}
   \caption{Linear stability analysis for $I=139$.}
   \label{fig:lin_stab}
\end{figure}
The shape of the curve is reported in Fig.~\ref{fig:lin_stab} for $I=139$.
This result confirms the need to go one step beyond the $1^{st}$ order analysis, to remove the
degeneracy of the Hamiltonian dynamics generated by the KGT equation.

The (in)stability is ruled by three terms: (i) the first one contributes to the stability of long wavelengths:
it originates from the term $\Gamma(1+I)g$ in Eq.~(\ref{eq:glin}), where it appears as a dissipative term in
the spatial evolution; (ii) the last term contributes to the stability of the short wavelengths: it
originates from the diffusive term in Eq.~(\ref{eq:qlin}); (iii)  the second term provides a
destabilizing wavelength-free contribution: it arises from the last two $\Gamma$ terms, which involve
the first derivative of $g$ and $f$, respectively.

The most unstable (least stable) mode is
\begin{equation}
k_{max} = (2I(1+I))^{1/4}~.
\end{equation}
For $I=139$ (the value typically considered in our simulations), $k_{max}=14.04\ldots$.
The corresponding eigenvalue is

\begin{equation}
\left(\frac{\lambda_R}{\Gamma}\right)_{max} = 3I -2\sqrt{2I(I+1)}
\end{equation}
It is positive if $I>I_\theta = 8$, a value which represents the laser threshold for space-time dynamics.
Recalling the relationship between $I$ and the pump 
$a = (I+1)(1-R)$, it corresponds to $a_\theta = 0.45$, to be compared with the original exact result $0.488\ldots$ (see e.g. ~\cite{Lugiato1986}).
This threshold value coincides also with the value determined by Perego et al. (see the appendix in ~\cite{Perego2020}). 
For large pump values, the instability is proportional to $I$, $\lambda_R/\Gamma \simeq (3-2\sqrt{2})I = (0.17\ldots)\times I$.

The leading term of the imaginary part is
\[
\lambda_I = \frac{2I}{k}~.
\]
It remains finite in the limit $\Gamma \to 0$, representing the frequency of the $k$th eigenmode.

Finally, notice that above threshold, the unstable modes cover a finite interval. For $I=139$,
$k\in [11.875:16.612]$ (see Fig.~\ref{fig:lin_stab})

\subsection{Numerical simulations}

The QHM  consists of a single ``dynamical" equation, Eq.~(\ref{eq:finalfs2}),
which determines the evolution of the electric field $f$. 
The population inversion $g$ is a ``slaved" variable: at any time $\tau$,
the spatial profile $g(\sigma,\tau)$ is obtained by integrating in space Eq.~(\ref{eq:finalgs}) for 
the given field $f(\sigma,\tau)$.
The equation for $g$ is dissipative and linear: $f$ plays the role of a periodic modulation.
The integration requires determining a periodic solution $g(\sigma,\tau)=g(\sigma-T,\tau)$.
Given the linearity of the equation, it follows that
$g_{fin} = v g_{in} + w$, where $g_{in}$ is the hypothetical initial condition (for $\sigma=0$)
while $g_{fin}$ is the final value reached for $\sigma=T$, while $v$ and $w$ are the (unknown) coefficients
resulting from the integration of the equation;
$v$ and $w$ can be determined by integrating Eq.~(\ref{eq:finalgs}) for two different choices of $g_{in}$ after
determining the two corresponding $g_{fin}$.
Once they are known, the periodic solution is obtained by imposing
$v g^* + w = g^*$, i.e. setting $g^* = w/(1-v)$.

\begin{figure}
\includegraphics[width=0.5\textwidth]{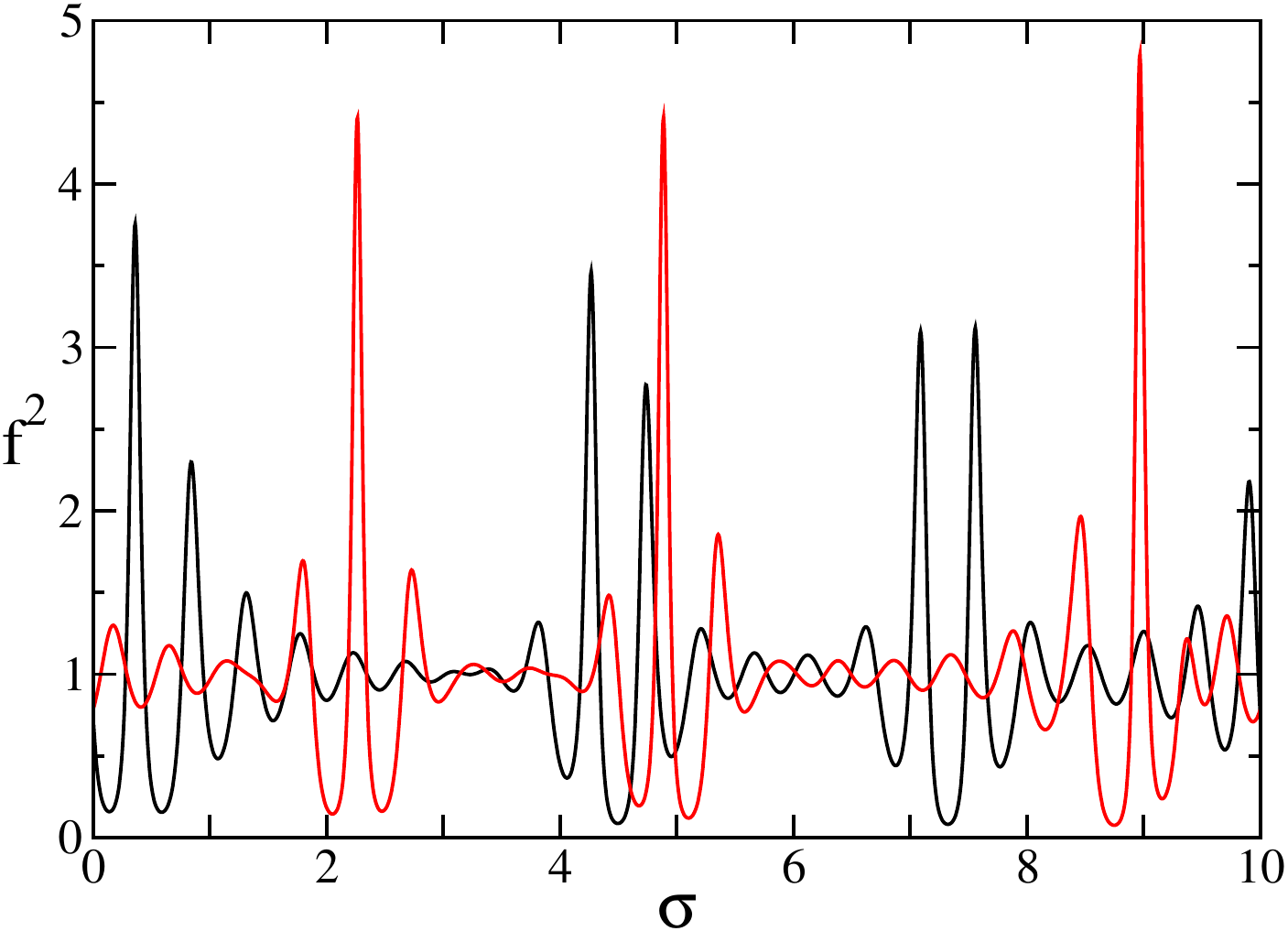}
   \caption{Two spatial profiles sampled during a simulation for $T=10$, $\Gamma = 10^{-3}$, and $I=139$.}
   \label{fig:profiles}
\end{figure}

Once $g(\sigma)$ is obtained, one can integrate the differential equation~(\ref{eq:finalfs2}) over time. 
We proceed by expanding $\partial_\tau f$ in Fourier modes, using a time step $\delta \tau$ typically of the order of $10^{-3}$.
$N=2048$ modes suffice for $\Gamma <\approx 10^{-4}$ (otherwise $N$ must be doubled and $\delta \tau$ lowered
possibly down to $10^{-4}$).
The resulting intensity profiles exhibit an approximately constant number $n$ of oscillations, typically
clustered in bursts of 3 to 5 peaks which form, propagate and annihilate along the $\tau$ axis.
Two exemplary profiles are reported in Fig.~\ref{fig:profiles}:
they have been obtained by integrating the QHM for $I=139$, $\Gamma= 10^{-3}$, and a length $T=10$.
The number of oscillations ($n \approx 21-22$) of oscillations
is consistent with the wavelength of the most unstable mode 
as from the linear stability analysis; indeed $n \simeq (Lk_{max})/(2\pi)$. 

\begin{figure}
\includegraphics[width=0.5\textwidth]{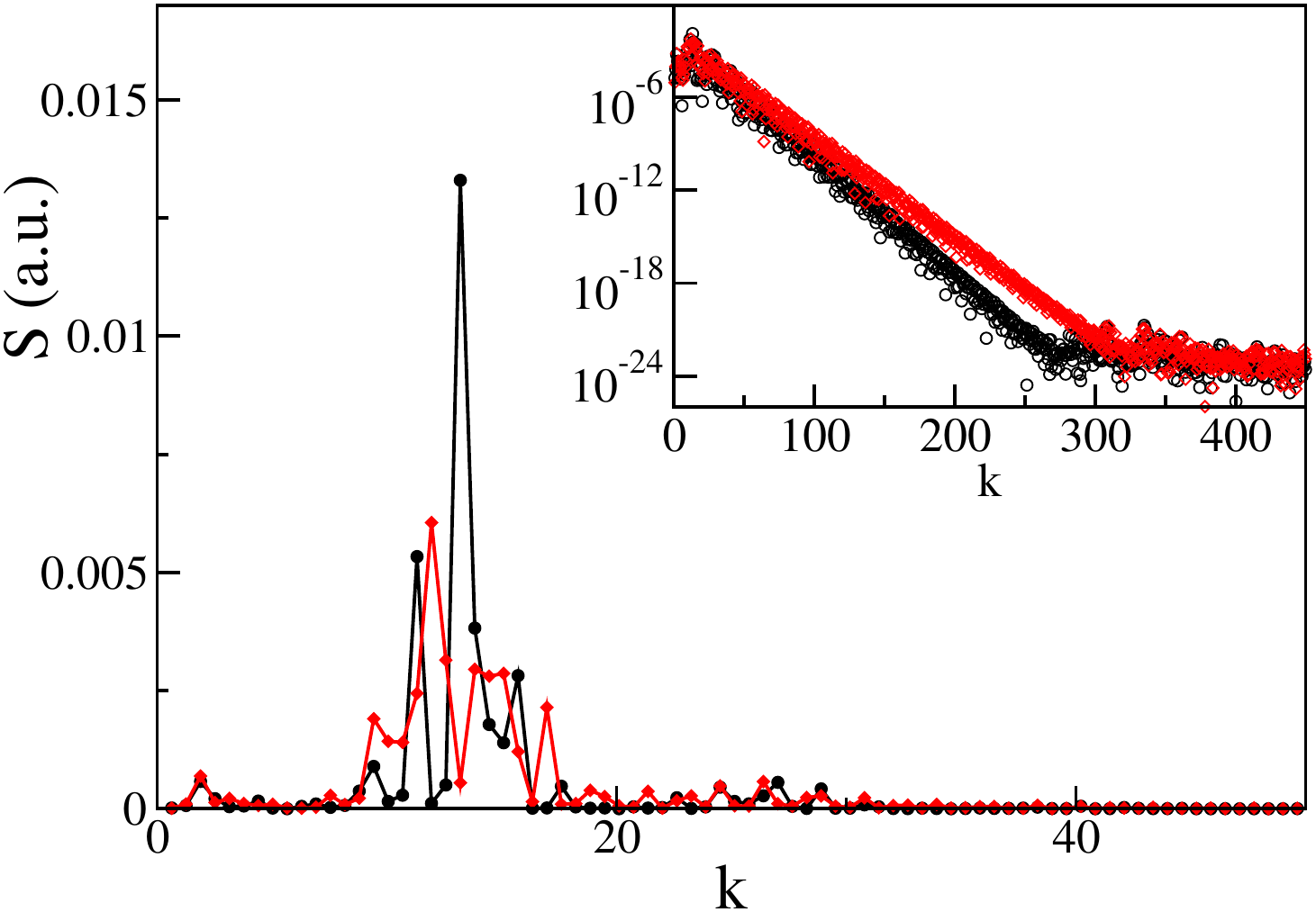}
   \caption{Power spectra of the two profiles in the previous figure.}
   \label{fig:spectra}
\end{figure}

A more quantitative representation is presented in Fig.~\ref{fig:spectra}, where the corresponding power spectra are plotted
in linear and semilogarithmic (see the inset) scales.
From the body of the figure it is clear that the relevant Fourier modes are those emerging from the linear stability analysis,
while the exponential decrease of the spectrum hints at the smoothness of the profiles. The final plateau above $k\approx 300$
is an unavoidable effect of the finite computational accuracy.

\begin{figure}
\includegraphics[width=0.5\textwidth]{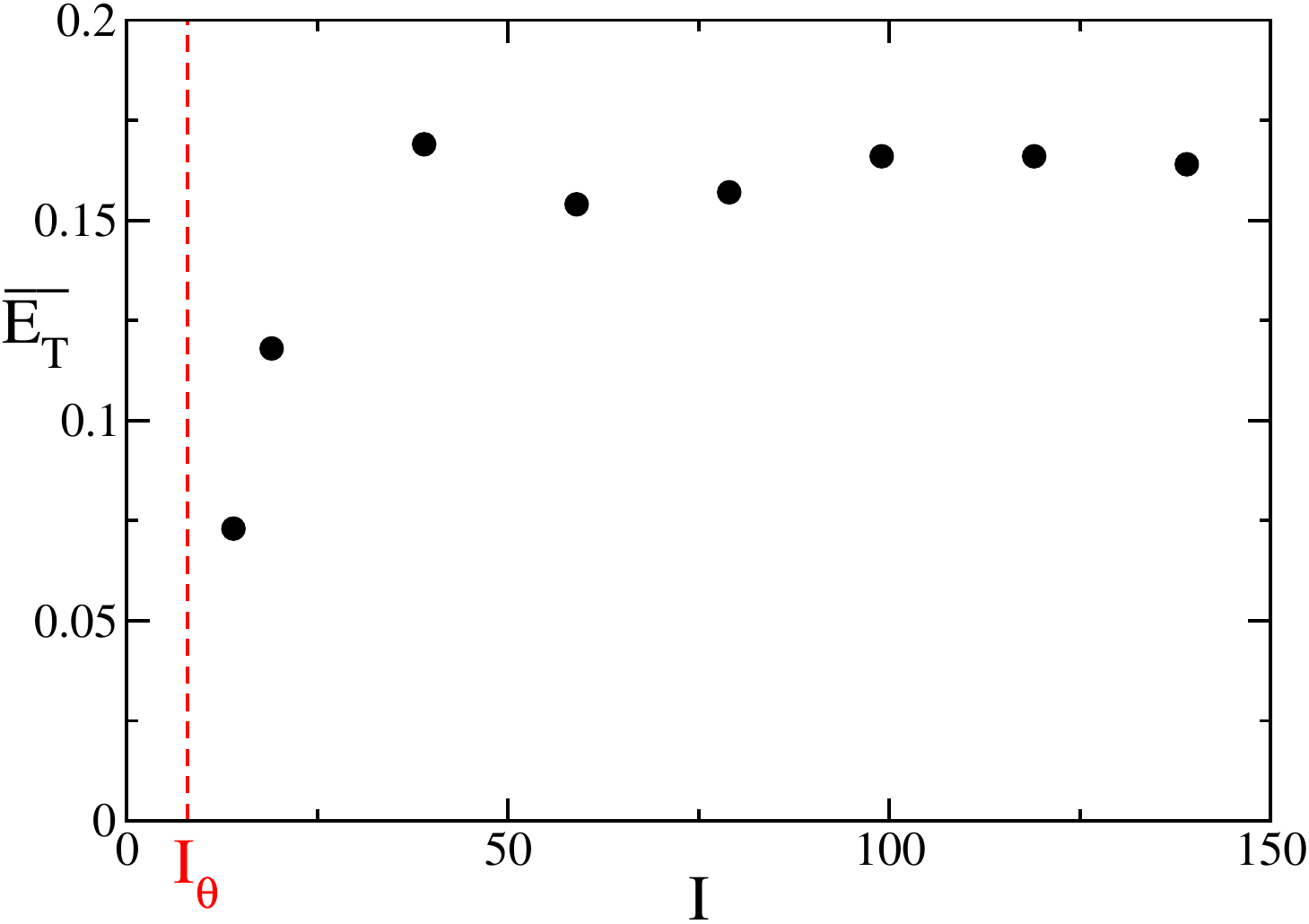}
   \caption{Dependence of the average energy on the effective pump $I$ for $\Gamma=10^{-4}$. 
   Simulations for different pump values have been made for different lengths $T$, in such a way
   that the number of unstable modes is kept constant; this means that for $I=14$, $T\approx 31.5$.}
   \label{fig:enerj0}
\end{figure}

We conclude whis section by discussing the dependence of the (time) average of the energy density $\overline{E_T}$ 
on the effective pump $I$.
The results of numerical simulations are reported in Fig.~\ref{fig:enerj0}: they have been obtained for
$\Gamma= 10^{-4}$.
The dashed vertical line at $I_\theta=8$ identifies the RNGH threshold, while $I=0$ denotes the first laser threshold, below
which the laser is not active. Both thresholds are independent of $\Gamma$.
Approching $I_\theta$ from above, the energy vanishes. This is obvious, since the field amplitude is constant below the
RNGH threshold. Upon increasing the pump, the energy rapidly reaches a saturation around 0.17. 
This awkward behavior is less surprising if we recall that 
the energy is defined with reference to the rescaled field intensity $\langle f^2\rangle \approx 1$.
Hence, since $E_T$ is, in the linear approximation, a quadratic function of $f$, we can conclude that it
is approximately proportional to the pump, if expressed in physical units.

The dependence of $\overline{E_T}$ on $\Gamma$ is a more subtle issue, that we address in Sec.~\ref{sec:energy}.

\section{The Coherent Master Equation} \label{sec:cme}
In this Section, we critically compare Eqs.~(\ref{eq:finalgs}, \ref{eq:finalfs2})
with the only model available in the literature, able, to our knowledge, to reproduce the RNGH threhsold.
Such a model was derived in Ref.~\cite{Perego2020}, by applying a second-order adiabatic 
elimination of the atomic polarization to the AB model.
It has been named Coherent Master Equation (CME) to stress the fact that it overcomes the shortcomings of the Haus model 
(known as the Master Equation in the field of mode locking systems).

Here, we discuss the CME in the free-running (ring) laser configuration (i.e. in the absence of intra-cavity devices).
Using our notations, the CME can be cast as
\begin{eqnarray}
\partial_\tau f &=& \left(  g    + \Gamma\partial_\sigma^2 \right) f  - \Gamma \partial_\sigma  
 \left( h f\right)\\
\Gamma \frac{(1-R)}{ R T } \partial_\tau  \langle g \rangle&=& I\left( 1- \langle f^2 \rangle \right) 
-\Gamma \langle g \rangle  -I\Gamma \langle  gf^2 \rangle \label{eq:gPrati}\\
\partial_\sigma h &=& I\left( \langle f^2 \rangle-f^2 \right)-
\Gamma h -\Gamma I gf^2 + 
\frac{\Gamma I}{2} \partial_\sigma f^2 +\Gamma I \langle gf^2\rangle ~,
\end{eqnarray} 
where $h = g-  \langle g \rangle$.

In practice, at any instant of time, given $f(\sigma)$ and the current
value of $\langle g \rangle$, one can
spatially integrate the equation for $h$ under the condition $\langle h \rangle = 0$. 
Once $h$ (and hence $g = \langle g \rangle + h$) is determined, the first
two equations can be then integrated in time.

At variance with the CME, $\langle g \rangle$ in the QHM
is implicitly obtained from the knowledge of the instantaneous field profile $f$,
instead of having to be integrated over time.
The corresponding equation is determined by integrating Eq.~(\ref{eq:finalg}) over space
\begin{equation}
- \Gamma  ( \langle g \rangle  + I \langle gf^2 \rangle) +
I(1-\langle f^2 \rangle) =0 \; .
\label{eq:gave_ae}
\end{equation}
This equation is equal to Eq.~(\ref{eq:gPrati}) of the CME,
except for the missing time derivative, a term that would be recovered if one averaged
the full equation Eq.~(\ref{eq:gsigma_full}) instead of Eq.~(\ref{eq:finalg}).
Within the framework of the QHM derivation, 
the absence of an explicit $\langle g \rangle$ dynamics is justified under 
the assumption of a large $R$ and not-too-small $T$.
It can be viewed as a standard adiabatic elimination.
On the other hand, whenever the $\partial_\tau \langle g\rangle$ term cannot be neglected,
similar terms neither included in the QHM  nor in the CME, 
should also be added to the equation for $f$.

\begin{figure}
\includegraphics[width=0.5\textwidth]{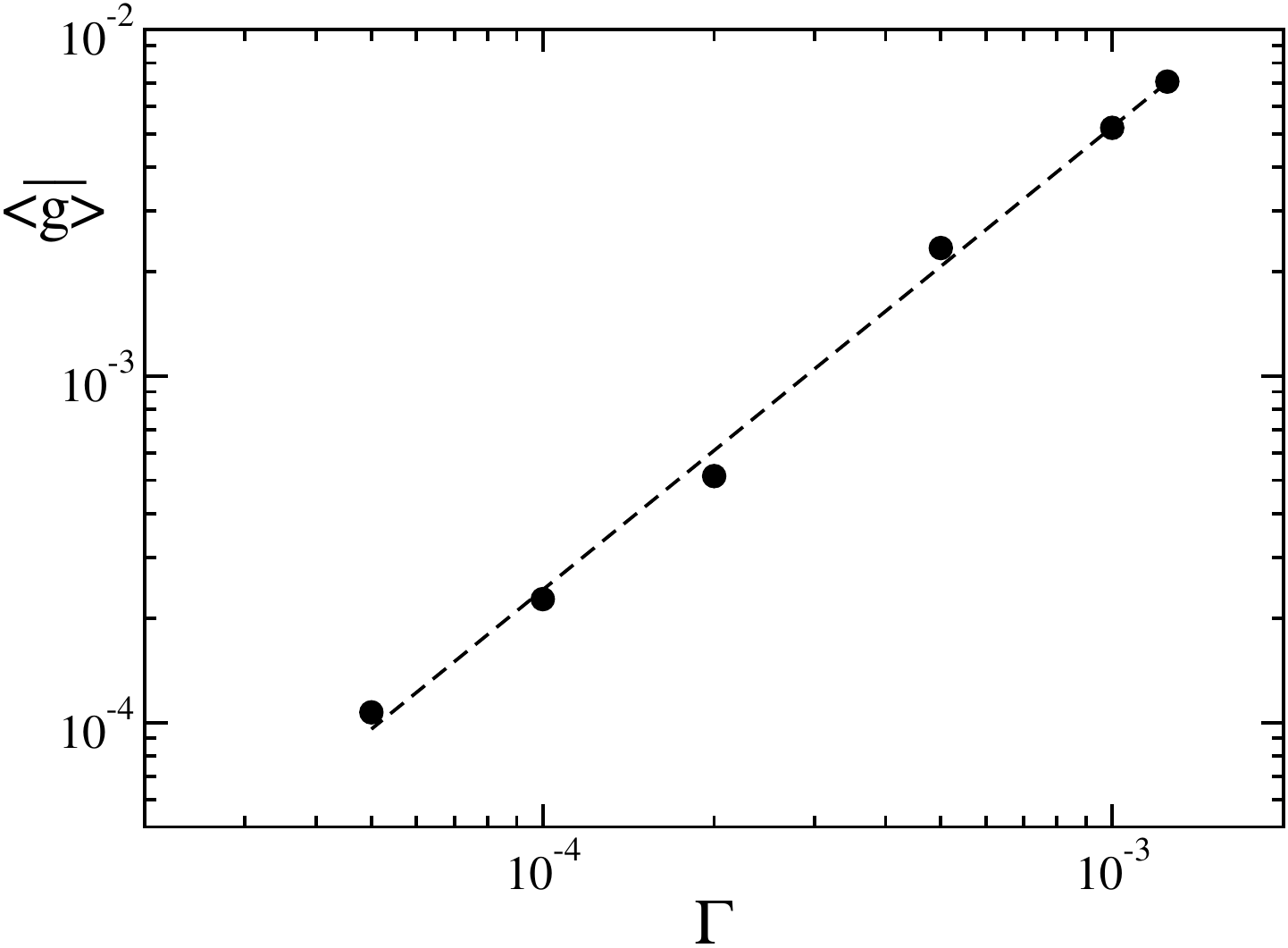}
\caption{Time average of $\langle g \rangle$ for different $\Gamma$ values. The slope of the power-law fit is 1.3}
\label{fig:g_ave}
\end{figure}

In order to complete the comparison, it is necessary to assess the amplitude of $\langle g \rangle$.
At leading order,
\begin{equation}
I \langle g f^2\rangle = \frac{I}{T}\int_0^L(gf^2)d\sigma \approx \frac{1}{T}\int_0^\sigma g \left(I -g_\sigma\right)d\xi =  
I \langle g \rangle  
\end{equation}
so that Eq.~(\ref{eq:gave_ae}) can be rewritten as
\begin{equation}
\langle g \rangle = 
\frac{I}{\Gamma(1+I)}
(1-\langle f^2 \rangle) \; . 
\end{equation}
This equation expresses the relationship between the size of $\langle g\rangle$ and the deviation of $\langle f^2 \rangle$ from 1.
In the Hamiltonian limit, both quantities vanish (being $\langle f^2\rangle$ a conserved quantity); 
for a small but finite $\Gamma$, we see that $\langle g \rangle$ is one
order lower in $\Gamma$ than $(1-\langle f^2\rangle)$.
However, scaling considerations do not suffice to establish the smallness of $\langle g \rangle$. 
Hence, we have performed several simulations, to determine the time average $\overline{\langle g \rangle}$
for different $\Gamma$-values. The results, reported
in Fig.~\ref{fig:g_ave}, suggest that $\langle g \rangle$ is even smaller than $\Gamma$ (a power law fit suggests a scaling
$\Gamma^\delta$ with $\delta \simeq 1.3$).
Consequently $(1-\langle f^2\rangle)$ is smaller than $\Gamma^2$.

By now returning to the first equation of the CME, we see that it is equivalent to Eq.~(\ref{eq:finalfs}), since
the replacement of $hf$ with $gf$ yields a correction at least of order $\Gamma^2$ (if not smaller) and is thus
irrelevant.
Similar arguments show that $h$ can be replaced by $g$ in the last equation and  $\langle f^2 \rangle$ by 1, generating
corrections of order $\Gamma^2$ again negligible. 

Altogether, we can conclude that the CME is essentially equivalent to the QHM, although slightly more complicated.
An important conceptual difference is that the absence of a perturbative representation in the CME, hides in general 
the closeness of the model to a Hamiltonian system and, in particular, the existence of a slow energy dynamics.

\section{Energy dynamics} \label{sec:energy}

At first-order the laser dynamics is described by the KGT model, which displays (in physical units):
(i) the ``fast" time scale $(\gamma_\perp\gamma_\parallel)^{-1/2}$ -- corresponding
to the period of the amplitude oscillations unstable above the 
RNGH threshold; (ii) the ``slow" time scale $\mathcal{T}/\Gamma$,
encoding the spatio-temporal Hamiltonian evolution.
Additionally, the KGT model exhibits two conserved quantities: $E_K$ and $E_P$.

This degeneracy is removed at the next order, by the perturbative terms of the QHM.
The energy dynamics is better described by referring to $s = \ln f$. 
In the appendix \ref{app:etot}, it is shown that $E_K$ and $E_P$ satisfy the following differential equations,
\begin{equation}
\frac{(E_K)_\tau}{\Gamma} =
-\frac{1}{2} \langle s_\sigma^2 \rangle  +\frac{7}{2}\langle {\rm e}^{2s}s_\sigma^2 \rangle -
(1+I)\langle s(e^{2s}-1) \rangle - 
\frac{1}{I} \langle s^2_{\sigma\sigma} \rangle ~,
\label{eq:ekindot_fina}
\end{equation}

\begin{equation}
\frac{(E_P)_\tau}{2\Gamma} = 
- \frac{1}{I} \langle  g^2 \rangle 
- \langle g^2 e^{2s}\rangle  
+\langle \mathrm{e}^{2s}(\mathrm{e}^{2s}-1) \rangle
-\langle \mathrm{e}^{2s}s_\sigma^2 \rangle
+ \langle g s_\sigma \rangle - \langle s_\sigma^2 \rangle   ~.
\label{eq:epotdot_fina}
\end{equation}
The two derivatives are both proportional to $\Gamma$, showing that the energy evolves
over the yet slower time scale $T/\Gamma^2$.

Altogether, the laser dynamics can be interpreted as a nonequilibrium steady state.
The KGT equation describes the underlying Hamiltonian system, whose statistical properties we are
interested in, while the terms proportional to $\Gamma$ of the QHM can be formally interpreted
as coupling terms with a fictitious external environment, which absorbs and releases energy.
The stationary regime is the result of a balance of the incoming and outgoing fluxes, 
represented by the terms on the r.h.s.'s. of Eqs.~(\ref{eq:ekindot_fina},\ref{eq:epotdot_fina}).

\begin{figure}
\includegraphics[width=0.5\textwidth]{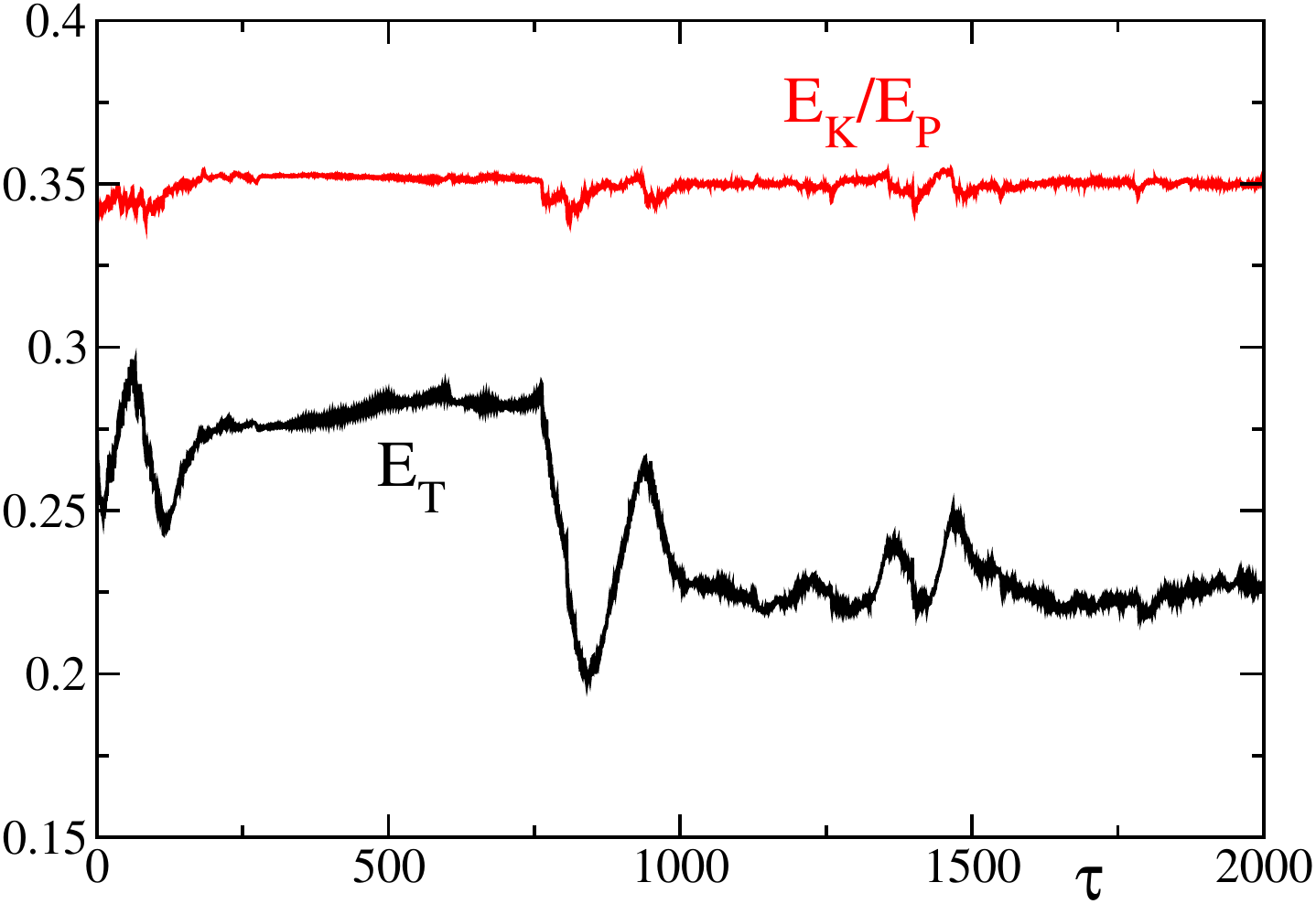}
\caption{Total energy density (black) and ratio between kinetic and potential energy (red), for $I=139$, 
$\Gamma = 1.25 \times 10^{-3}$ and $T=10$.}
\label{fig:energies}
\end{figure}

The two energy equations are not closed and even involve collective observables that are not conserved at
first-order. Hence, they cannot be directly used to determine the energy dynamics.
We can nevertheless proceed indirectly by simulating the full QHM.
In Fig.~\ref{fig:energies}, we report the time trace of the total energy density $E_T$ and of the ratio $\rho = E_K/E_P$, for
$\Gamma = 1.25\times 10^{-3}$.
The energy density exhibits significant fluctuations, analogously to the behavior reported
in~\cite{Politi2023}, where the original delayed model was integrated for the same parameter values.
The jumps are a consequence of a changement in the number of bursts along the spatial profile
(see Fig.~\ref{fig:profiles}).
On a more quantative level, the energy values are slightly smaller than those displayed by the delayed laser model
(in which case the energy fluctuates in the range range $[0.28,0.36]$~\cite{Politi2023}).
The discrepancy is quite likely due to higher order corrections not included in the QHM.
On the other hand, the ratio $\rho$ exhibits, instead, small fluctuations around 0.35, in full agreement with the
original laser model.

\begin{figure}
\includegraphics[width=0.5\textwidth]{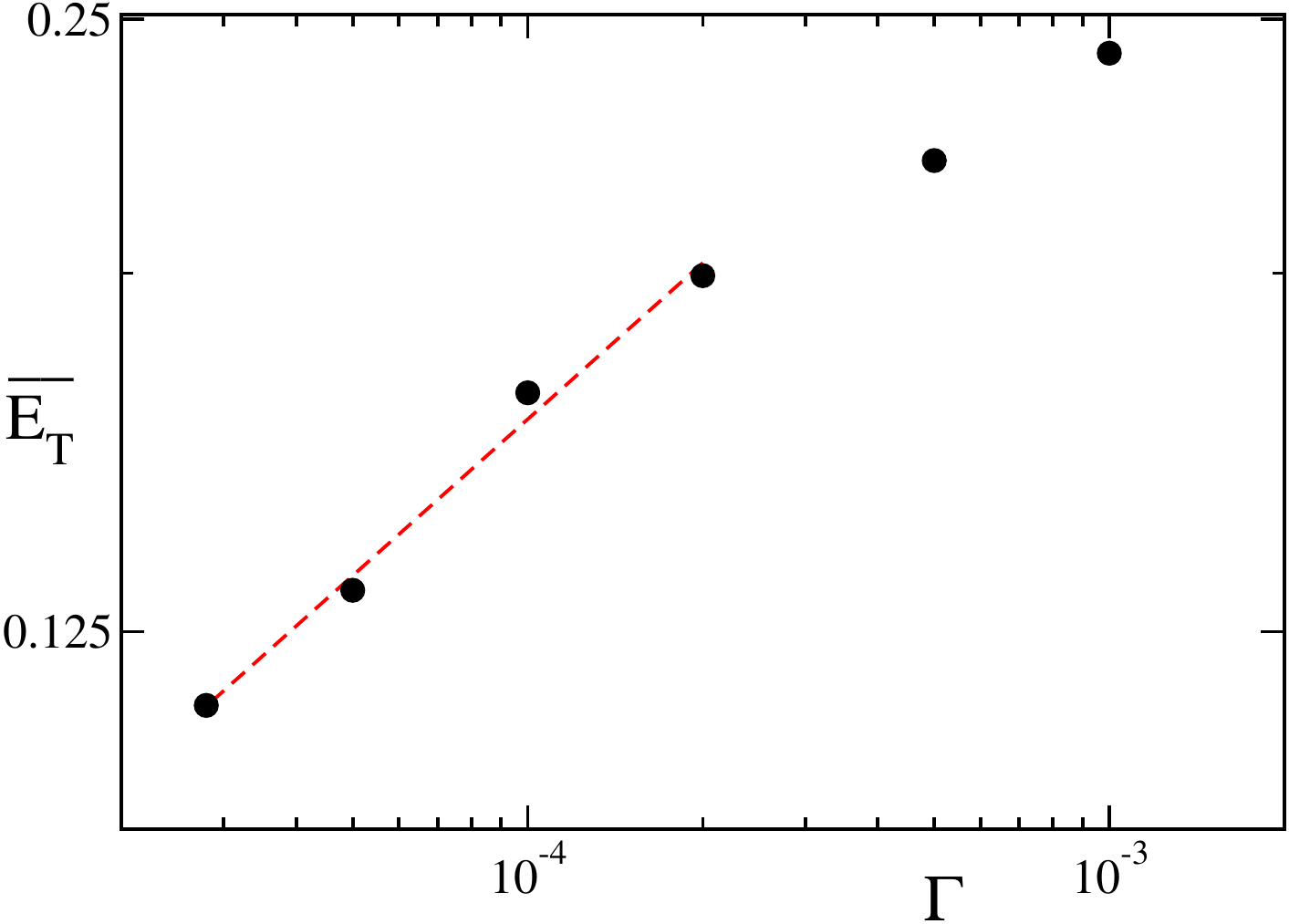}
\caption{Average total energy vs. $\Gamma$ in log-log scales. The dashed line corresponds to a slope 0.25.}
\label{fig:ener_gamma}
\end{figure}

In Fig.~\ref{fig:ener_gamma} we report the average total energy density $\overline{ E_T }$ numerically 
determined for a series of $\Gamma$ values down to $\Gamma =2.8\times 10^{-5}$, a realistic value for Erbium-doped fiber lasers.
All data are affected by large statistical fluctuations, difficult to quantify.
Anyway, it is clear that $E_T$ decreases with $\Gamma$, possibly following a power law $\Gamma^\eta$.
Without pretending to be too reliable, we find $\eta \approx 0.25$.

\begin{figure}
\includegraphics[width=0.45\textwidth]{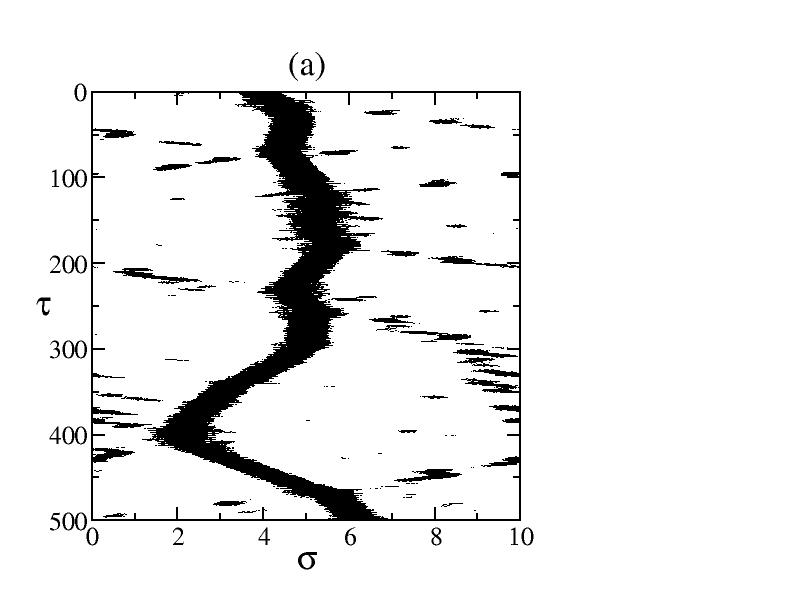}
\includegraphics[width=0.45\textwidth]{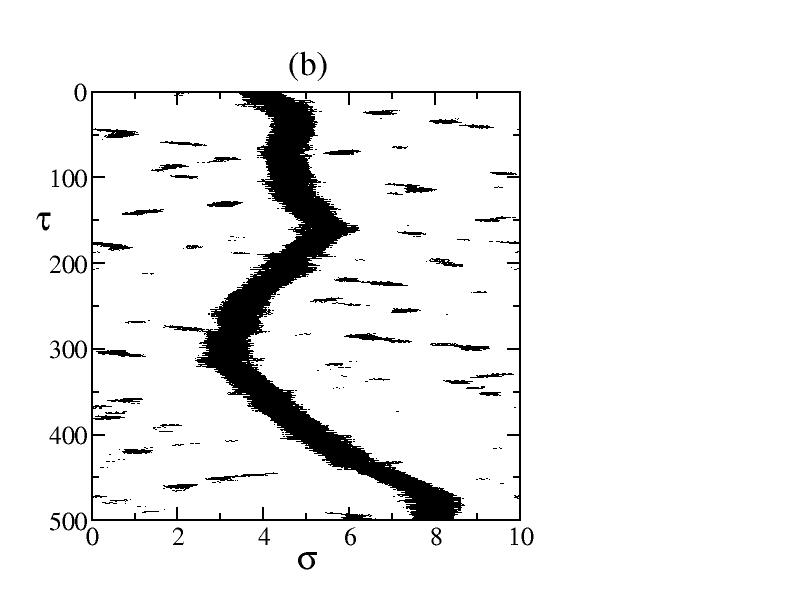}
   \caption{(a) Pattern sampled from a simulation of the QHM for $\Gamma = 2.8\times 10^{-5}$; 
(b) pattern generated by integrating the KGT model, starting from the same initial condition as in (a). 
Only points where the intensity is larger than 1.7 are reported.
In both cases the selected frame moves with velocity 2.42.}
   \label{fig:patt1}
\end{figure}

In order to test, at least qualitatively, the relevance of the perturbative terms for the smallest meaningful $\Gamma$ value, we
have compared the pattern obtained by integrating the QHM, with that obtained by integrating the KGT model, starting
both from the same initial conditions. The results are presented in Fig.~\ref{fig:patt1}. The effect of the
$\Gamma$ terms becomes visible on a time scale larger than $\tau \approx 100$.

\subsection{Minimal model}

\begin{figure}
\includegraphics[width=0.5\textwidth]{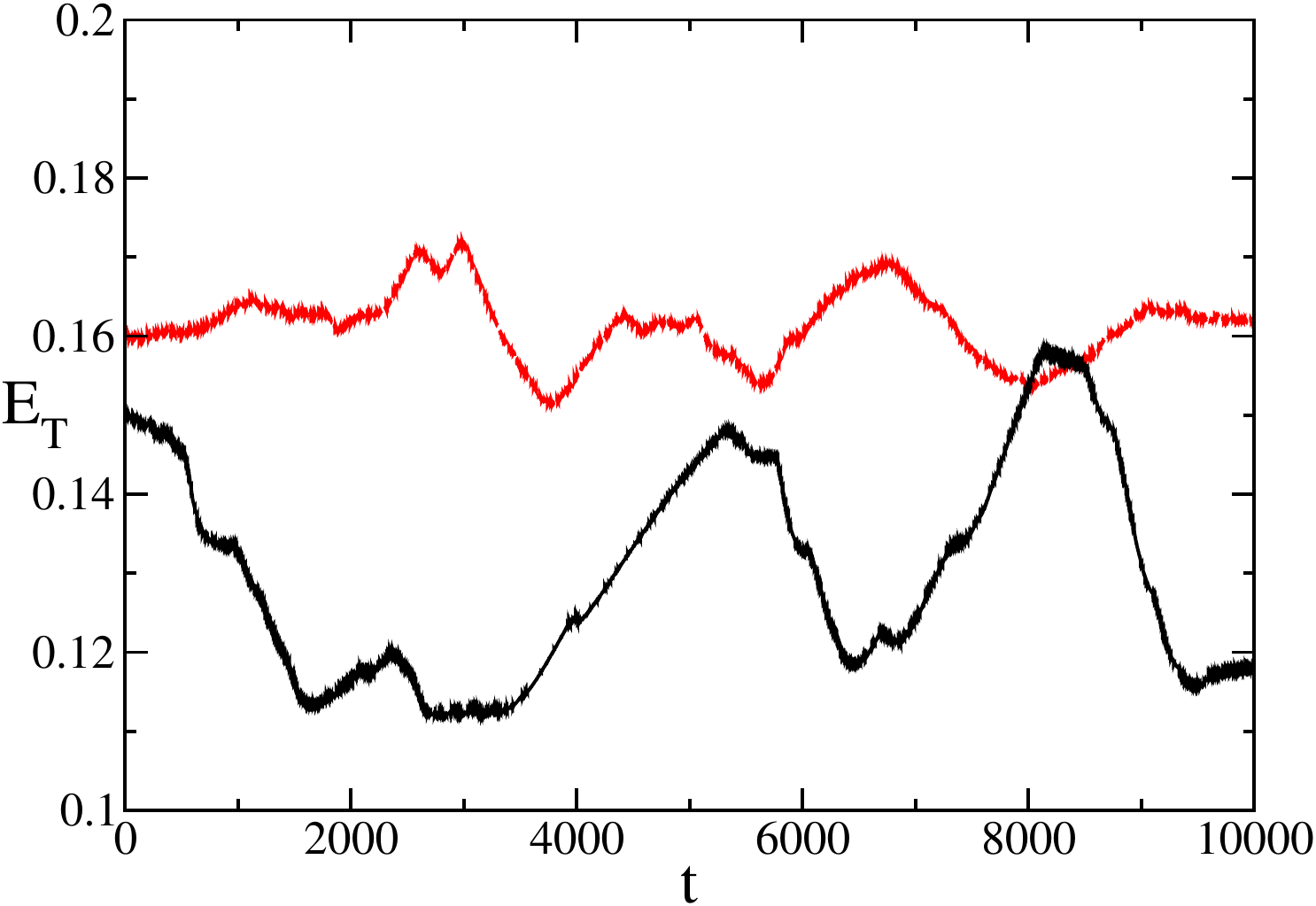}
\caption{Total energy for $\Gamma=5\times 10^{-5}$ as from the simulation of the QHM (black solid line) and the MM 
(red dashed line).}
\label{fig:ener_mm}
\end{figure}

An important question is still open: how is the $E_T$-value selected by the dynamics? 
Here, we develop some scaling arguments based on a further simplification of the QHM.
The data reported in Fig.~\ref{fig:ener_gamma} show that $E_T$ is small for realistic $\Gamma$ values; this means that
also the (relative) field fluctuations are small (the zero-energy state corresponds to the stationary regime $f=1$).
Actually, in the QHM (Eqs.~(\ref{eq:finalgs},\ref{eq:finalfs2}))
there are two types of nonlinear terms: (i) those of leading order, responsible for the nonlinear Toda force within the 
Hamiltonian dynamics; (ii) those proportional to $\Gamma$. 
It is logical to conjecture that the latter ones are negligible, being at the same time much smaller than the former ones
(because of the proportionality to $\Gamma$) and smaller than their linear perturbative counterparts.
Within this approximation, $gf$ can be replaced by $g$ and $gf^2$ by $g$ (in practice, setting $f=1$), obtaining
\begin{equation}
f_\tau = gf -\Gamma g_\sigma + \Gamma f_{\sigma\sigma} 
\end{equation}
\begin{equation}
g_\sigma = I(1-f^2) - \Gamma \left ( g(1+I)   -I f_\sigma \right )
\end{equation}
The quality of this approximation has been checked comparing this minimal model (MM) with the QHM
for $\Gamma= 5.\times 10^{-5}$. In Fig.~\ref{fig:ener_mm}, we see that the energy of the MM is slightly larger than that of
the QHM (the average $E_T$ is around 0.16 instead of 0.13). Another difference are the larger fluctuations exhibited
by the QHM.
Since we are not interested in a fully quantitative agreement, but simply in understanding the selection mechanisms
of a specific average energy, we feel authorized to proceed with the discussion of the MM.

Let us consider a hypothetical initial condition such that nonlinearities are negligible. So long as this is true, 
the evolution can be decomposed into the independent dynamics of the Fourier eigenmodes. 
Hence, according to the stability analysis presented in Sec.~\ref{sec:linestab}, the amplitude
of those modes with a wavenumber inside the unstable range would diverge, while all the others would decrease
exponentially to zero.
According to linear stability analysis, we can therefore approximately state that the perturbative linear terms
of the MM contribute to an incoming energy flux towards the unstable modes
$\Phi_u \approx \alpha_u \Gamma E_u$, proportional to $\Gamma$ and to the energy $E_u$ contained in
such modes. Similarly, we expect an outgoing flux $E_s \approx \alpha_s \Gamma E_s$ from the stable
modes.

\begin{figure}
\includegraphics[width=0.45\textwidth]{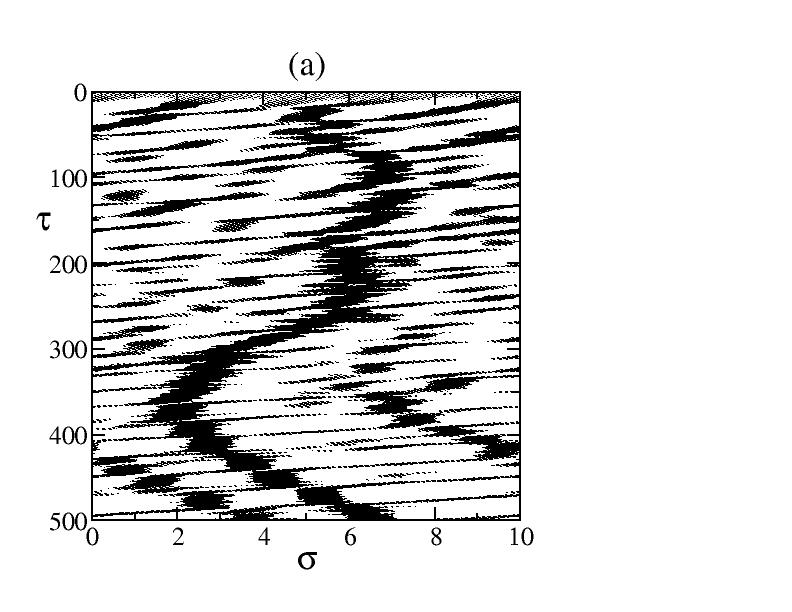}
\includegraphics[width=0.45\textwidth]{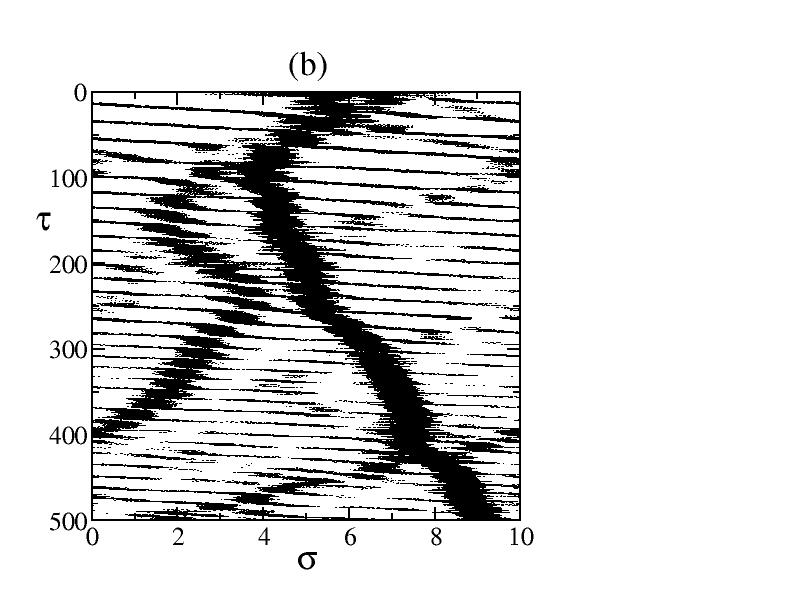}
\caption{Two patterns obtained by integrating the KGT model starting from an initial condition of the type
$s = a_1\cos(2\pi k_1 \sigma /T ) + a_2\cos(2\pi k_2 \sigma /T ) -b $.
In panel (a): $a=0.29$, $k_1 = 22$, $a_2=0$ and $b=0.04119$;
in panel (b) (b) $a=0.26$, $k_1 = 22$, $a_2=0.13$, $k_2=21$ and $b=0.04166$. The offset $b$ is chosen so as to
ensure $\langle f^2 \rangle =1 $. Only points where the intensity is larger than 1.7 are plotted; the frame moves
with velocity 2.55 in both panels.}
\label{fig:patt2}
\end{figure}

However, at some point, nonlinearites enter the game, transferring energy among the various modes.
This phenomenon is qualitatively illustrated in Fig.~\ref{fig:patt2}, where we plot the evolution starting
from a unimodal (panel (a))  and a bimodal (panel (b)) state~\footnote{Unimodal and bimodal refer to the
$s = \ln f$ representation}.

In both cases, kinetic and potential energy have been selected to be very close to those of the simulations in
Fig.~\ref{fig:patt1}. In both cases, there is a relatively fast transfer to other modes, but the
observed patterns are different between themselves and different from that of the QHM, indicating 
the (expected) presence of yet longer time scales, due to weak nonlinear mechaninsms, i.e. to the
quasi integrability.

The flux $\Phi_{us}$ from unstable towards stable modes is the last ingredient, needed to determine the properties
of a stationary nonequilibrium state: it indeed acts as a bridge compensating the  
above mentioned fluxes exchanged with the ``external" environment.

It is natural to assume that $\Phi_{us}$ depends mostly on $E_u$ (the source of energy): 
we assume $\Phi_{us} \approx E_u^\beta$ with $\beta>1$.
A stationary state exists if and only if the energy fluxes balance each other in both families of modes. 
With reference to the unstable modes, $\alpha_u \Gamma E_u \approx E_u^\beta$, 
so that $E_u \approx \Gamma^{1/(\beta-1)}$.
Compensation for the stable modes requires that $E_s$ scales in the same way $E_s \approx \Gamma^{1/(\beta-1)}$.
Hence, also the total energy $E_T$ scales in the same way. 
Since the intramode flux is determined by nonlinearities of the KGT, we expect $\beta>1$, which means that 
$E_T$ is expected to decrease with $\Gamma$, as indeed seen from the direct simulations reported in Fig.~\ref{fig:ener_gamma}.
From the empirical scaling therein observed -- $E_T \approx \Gamma^{0.25}$ --  we can conjecture that
$\beta=5$, but this value should be taken with tongs, given the sketchiness of the argument.
The important point is that the average total energy decreases with $\Gamma$, because this is the only way for
the internal nonlinear fluxes to compensate the decreasing flux (proportional to $\Gamma$) with the external environment.

\subsection{Ratio of energies}

So far, we focused on the total energy, without distinguishing $E_K$ from $E_P$.
Numerical simulations, performed for different $\Gamma$- and pump-values
consistently supply $\rho$ values around 0.35.

Here below we provide a simple justification of this numerical result.
From the stability analysis, it turns out that the range of unstable modes is rather narrow.
For $I=139$, we have seen that $k = 2\pi n/T \in [11.87,16.61]$; if
$T=10$, as selected in our simulations, the unstable modes are those such that the integer $n$ belongs 
to the interval $[19, 26]$.
This is consistent with the number and wavelength of active modes revealed by direct numerical simulations
as seen in Fig.~\ref{fig:spectra}. 

By taking the narrowness to the extreme, we assume a strictly unimodal state

\begin{equation}
s(\sigma) = A\cos k\sigma ~,
\end{equation}
where $k$ is a yet unspecified  wavenumber.
The ratio of kinetic to potential energy is
\begin{equation}
\rho = \frac{E_K}{E_P} = \frac{A^2k^2}{4I [ \langle {\rm e}^{2A\cos k\sigma} \rangle -1] }
\end{equation}
In the linear limit ($A\ll 1$)
\begin{equation}
\rho = \frac{E_K}{E_P} = \frac{k^2 }{4I} 
\end{equation}
With reference to the most unstable  wavenumber $k_{max}$
\[
\rho \simeq  \frac{1}{4} \sqrt{2(1+1/I)}~.
\]
For $I=139$ (the value chosen in our simulations), $\rho = 0.354$, extremely close to the value 
observed in direct simulations (see Fig.~\ref{fig:energies}).
This explains the origin of the spontaneous selection of $\rho$: it corresponds to the
ratio observed for the most unstable spatial frequencies in the small amplitude limit.

\section{Conclusions and open problems}

In this paper we have derived and analysed the quasi-Hamiltonian model (QHM), a set of two PDEs that can
be effectively used to perform realistic
numerical simulations of ring-laser dynamics, in the presence of a fast relaxation of the atomic polarization.
It turns out that the overall dynamics is nearly Hamiltonian.
Accordingly, extremely long time scales are naturally present, emerging from the slow evolution of two global quasi conserved
quantities (the ``kinetic' and  ``potential"  energy). 
The two pseudo-energies are self determined by the second-order correction terms; interestingly,
in the limit $\Gamma \to 0$, the energies become progressively smaller, meaning that the dynamics
is increasingly integrable (linear).
As a result, an additional source of complexity arises due to the exceptionally long thermalization
times.
In the attempt to capture the relevant features of the laser dynamics, we propose a simplification of the QHM 
-- we call it the minimal model -- which helps to understand how nonlinearities contribute to the establishment
of a stationary regime.
Still the mechanisms responsible for the relevant energy variations observed in long simulations are unclear.
We have noticed they are associated with structural changes of the intensity profiles (number of ``bursts"),
but a less phenomenological analysis is surely required to make further progress.

Moreover, there is still much to do from a mathematical point of view. Can one recast the adiabatic elimination
of the field $U$ within the general framework of an inifinite-dimensional center (inertial) manifold?
Can one reformulate our space-time representation in rigorous terms by implementing a multiple-scale technique?
These tasks are probably quite hard since the starting model is a neutral delayed equation:
i.e. it belongs to a largely unexplored class of dynamical systems.
Furthermore, the final QHM itself is quite peculiar; it formally involves two fields, but the
population inversion does not have its own dynamics. It is a pseudo mean-field which induces
long-range interactions via the integration along the spatial direction. 

Experimental tests are more than welcome to validate our theoretical predictions.
Erbium-doped fiber lasers are perhaps the most appropriate experimental testing ground 
since there $\Gamma \simeq 10^{-4}$ if not smaller.
First of all, it would be worth finding evidence of the extremely slow evolution of 
coherent structures in the pseudo spatio-temporal representation associated to the energy dynamics.
It should be, however, noticed that the nature of the QHM implies that the underlying evolution
is highly sensitive to perturbations (being them either stochastic or deterministic), because they may 
easily induce significant variations of the pseudo-energies. 
Real fiber lasers are surely affected by various kinds of noise,
not to mention possible perturbations due to non perfect resonance, or propagation losses, herein neglected.
We plan to address these questions in a future work.

Furthermore, it will be worth to make additional theoretical efforts to extend the
perturbative approach (based on the assumption of $\Gamma \ll 1$) to the off-resonance case. 
Long-ago, it was found that the center manifold technique, developed to unravel the quasi-Toda behavior of 
single-mode resonant class-B lasers~\cite{Oppo1986},
can be successfully extended out-of-resonance \cite{Oppo89}. We do not see conceptual reasons why this robustness should not hold
in the general multi-mode case.

A last enlightening venue to explore is the laser behavior in the presence of intra-cavity devices,
a setup where e.g. the Haus equation has proven effective. This would help to clarify the role of the pseudo-energies
in a context where experiments are more easily doable and to better understand the differences with a model
which, instead fails to reproduce the free-running dynamics.

\begin{acknowledgments}
\end{acknowledgments}
We warmly thank S. Yanchuk for several discussions on this highly complex setup.

\newpage 
\appendix
\section{From AB to delayed model}
\label{app0}

The Arecchi-Bonifacio equations are the reference model, introduced in 1965 to describe the coherent amplification of electric field in a two-level medium and, later on, broadly used to study the onset of longitudinal instabilities 
in a ring laser with unidirectional propagation.
Here, we consider this setup restricting ourselves to the resonant case, starting from the assumption that
the polarization $\mathcal{P}$ and the electric field $\mathcal{F}$ are real. 
With reference to a comoving frame, the model can be written as ~\cite{Giacomelli2021},
\begin{eqnarray}
\frac{\partial\mathcal{F}}{\partial y}  &=& \frac{a}{2}\mathcal{P},  \nonumber \\
\frac{\partial\mathcal{P}}{\partial \hat t} &=& \mathcal{D}\mathcal{F}-\mathcal{P}, \label{MB}\\
\frac{\partial \mathcal{D}}{\partial \hat t} &=& \gamma \left [1-\mathcal{D} - \mathcal{F}\mathcal{P})\right ]~, \nonumber
\end{eqnarray}
where $\mathcal{D}$ is the population inversion, while
$y$ is a suitably scaled spatial variable ($y\in[-1,1]$);
$\gamma$ is the ratio $\gamma_\parallel/\gamma_\perp$ between the population and the polarization decay rate, respectively;
the time $\hat{t}$ is expressed in units of $\gamma_\perp^{-1}$, while
$a$ is the pump parameter, controlling the energy flux injected into the laser. 
The boundary condition is $\mathcal{F}(y=-1,t) = R \mathcal{F}(y=1,t-\mathcal{T})$,
$R$ being the mirror reflectivity and $\mathcal{T}$ the the round-trip time across the cavity.

In Refs.~\cite{Giacomelli2021,Politi2023}, it was shown that
this set of equations is well approximated by a spaceless delayed model, here presented in the most compact form,
\begin{eqnarray}
F &=& F^d+ \Gamma \frac{1-R}{R}U \label{orig01} \\
 \dot U &=& \frac{R}{\Gamma} [-U + GF - \dot F^d] \label{orig02} \\
\dot G &=& -\Gamma G + I(1- F^2 -\Gamma FU) \label{orig03}  \; .
\end{eqnarray}
Here, $\Gamma = \sqrt{\gamma}$ will play the key role of a smallness parameter, used in the perturbative expansion discussed
in the body of the paper.
The new time variable is $t = \Gamma \hat t$, so that the delay is $T = \Gamma \mathcal{T}$.
The pump parameter is also rescaled to the system losses, introducing the "effective" pump  as
\begin{equation}
	I = \frac{a}{1-R} -1  \; .
\label{eq:pump}
\end{equation}

Hence, the field is $F(t) = \mathcal{F}(1,\hat t)/\sqrt{I}$, scaled in such a way that the stationary value is equal to 1
(consequently $F^d(t) \equiv F(t-T)$).
The population inversion has been redifined as
$G = (\langle \mathcal{D}\rangle(I+1) -1)/\Gamma$: in practice, it is equal to zero in the stationary regime, while the rescaling
by $\Gamma$ is useful to deal with a variable of order 1, in the small-$\Gamma$ limit.
Finally, the most tricky change of variable is

\begin{equation}
U = (\langle \mathcal{P} \rangle (I+1)-\mathcal{F}(1,\hat t))/(\Gamma \sqrt{I})
\end{equation}
justified by the fact its zero-order adiabatic elimination suffices to obtain a meaningul set of equations,
even though it is still insufficient to describe the laser dynamics. In fact, this is the starting point of this work.

\section{$2^{nd}$-order perturbative expansion}
\label{app1}

The next order in the adiabatic elimination of $U$ yields (see Eq.~(\ref{orig2}))
\begin{equation}
U_2 =  U_1 - \frac{\Gamma}{R} \dot U_1~,  \label{ae2}
\end{equation}
With the help of Eq.~(\ref{ae1}), Eq.~(\ref{ae2}) can be rewritten as
\begin{equation}
U_2 =  G_2F_2 -\dot F_2^d 
 -\frac{\varepsilon}{1-R} [ \dot G_2F_2 + G_2\dot F_2 -\ddot F_2^d)]
\end{equation}
It is useful, for later convenience, to eliminate the delayed functions in the r.h.s.. 
This can be done by approximating $\ddot F_2^d$, at leading order, with $\ddot F_2$, while
the replacement of $\dot F_2^d$ requires including first-order corrections. Altogether,
\begin{equation}
\dot F_2^d \approx \dot F_2-\varepsilon \left(\dot G_2 F_2 +\dot F_2 G_2- \ddot{F}_2 \right) 
\end{equation}
As a result
\begin{equation}
U_2 =  G_2F_2 -\dot F_2
- \frac{\varepsilon R}{1-R} [ \dot G_2F_2 + G_2\dot F_2 -\ddot F_2  ]
\end{equation}
Finally, it is legitimate to approximate
$\dot G_2$ with the $\dot G_1$-expression of Eq.~(\ref{ord1b}), obtaining
\begin{equation}
U_2 =  G_2F_2 -\dot F_2 
 - \frac{\varepsilon R}{1-R}[ I(1- F_2^2) F_2 + G_2\dot F_2 -\ddot F_2  ]~.
\end{equation}

We can now replace this $U_2$ expression in the field equation (\ref{orig1}),
obtaining
\begin{eqnarray}
F_2 &=&  F_2^d + \varepsilon  \left( G_2F_2 - \dot F_2\right)- \label{eq:2ndF} \\
&&
\frac{\varepsilon^2 R}{1-R} \left( I(1- F_2^2)F_2 + G_2\dot F_2- \ddot F_2\right)  \nonumber
\end{eqnarray}
Eq.~(\ref{orig3}) is easier to process as we do not need to include the term of order $\varepsilon$ in the $U$ expression.
Proceeding as above, we obtain
\begin{equation}
\dot G_2 = I((1-F_2^2) - \frac{\varepsilon R}{1-R} \left ( G_2 + I F_2(G_2F_2 -\dot F_2 ) \right )
\label{eq:2ndG}
\end{equation}

Eqs.~(\ref{eq:2ndF},\ref{eq:2ndG}) represent the full model at second order.

As for the first order, it is convenient to introduce a space-time representation.
Moreover, from now on, we drop the subscript for the sake of simplicity and use lower-case letters.

With the help of Eqs.~(\ref{st1},\ref{st2}). Eq.~(\ref{eq:2ndF}) can be rewritten as
\begin{eqnarray}
&& \partial_\tau f \left (1+ \frac{\varepsilon}{T} \right )  + \partial_\sigma f  = \\
&& = gf + \frac{\varepsilon}{2}\partial_\tau^2 f -
\frac{\varepsilon R}{1-R} \left( I(1- f^2)f + g\partial_\sigma f - \partial_\sigma^2 f \right)  \nonumber
\end{eqnarray}

The perturbative terms formally enlarge the phase-space dimensionality, as they include a second-order time derivative.
However, thanks to their perturbative nature, one can remove the higher-order derivatives, expressing them as approximate
functions of the field $f$ (and $g$).

With the help of Eq.~(\ref{approx1}), we can approximate
\begin{equation}
\partial_\tau^2 f \approx - \partial_{\sigma\tau}f + f \partial_\tau g  + g \partial_\tau f \label{ftautau}
\end{equation}
Now, again with the help of Eq.~(\ref{approx1}), we can eliminate the time derivatives in the first and last term in the r.h.s
\begin{equation}
\partial_{\sigma\tau}f \approx -\partial_\sigma^2f + f \partial_\sigma g +g \partial_\sigma g, \qquad 
\partial_{\tau}f \approx -\partial_{\sigma}f + g f 
\end{equation}
so that
\begin{equation}
\partial_\tau^2f \approx 
\partial_\sigma^2f - f \partial_\sigma g -2g\partial_\sigma f + g^2 f + f \partial_\tau g  \label{eq:ftau2}
\end{equation}
More elaborate is the perturbative treatment to get rid of $\partial_\tau g$.
From the time-derivative of Eq.~(\ref{approx2}) (i.e. neglecting $\varepsilon$ terms), we find that 
\begin{equation}
\partial_{\tau\sigma}g \approx -2 I f \partial_\tau f
\end{equation}
Still in the same approximation, we can integrate in space this equation, obtaining
\begin{eqnarray}
\partial_\tau g /I &\approx & -2 \int  d\xi f \partial_\tau f  + C_2  \label{eq:gtau} \\
& \approx & -2 \int  d\xi f \left( -\partial_\sigma f + g f \right) + C_2 \approx 
f^2 -2 \widehat{g f^2} + C_2  \nonumber
\end{eqnarray}
where $C_2$ represents an integration constant to be suitably determined.
It can be fixed by considering the spatial average of the above equation. 
At zero order, we know that $\langle g \rangle = 0$.
Therefore, at the same order of approximation, its time derivative vanishes so that
\begin{equation}
0 = \langle f^2\rangle -2\langle \widehat{g f^2} \rangle +C_2
\label{eq:cond}
\end{equation}
and hence,
\begin{equation}
C_2 = 2\langle \widehat{g f^2} \rangle - 1
\label{eq:cond0}
\end{equation}
Practically, once a generic integral $\widehat{gf^2}$ has been determined, this equation yields the proper $C_2$ value.

Altogether, Eq.~(\ref{eq:ftau2}) can be written as
\begin{equation}
\partial_\tau^2 \approx g^2f -2g\partial_\sigma f -f\partial_\sigma g + \partial_\sigma^2 +
I f \left(  f^2 -2 \widehat{g f^2} + C_2 \right )
\end{equation}
Hence, the field equation of the second-order model can be written as
\begin{eqnarray}
&& \partial_\tau f\left (1+ \frac{\varepsilon}{T} \right )  + \partial_\sigma f = gf - \nonumber \\
&& \frac{\varepsilon}{1-R} \left [
RIf - \frac{1+R}{2} \left ( If^3  + \partial_\sigma^2 f\right) + g \partial_\sigma f\right ] 
\label{eq:finalf} \\
&& + \frac{\varepsilon}{2}
[g^2f - f\partial_\sigma g -  2 I f \widehat{gf^2}+IfC_2 ]  \nonumber
\end{eqnarray}

Now, we process the $g$ equation in a similar way. The original Eq.~(\ref{orig3}) can be written as
\begin{equation}
\partial_\sigma g -I(1-f^2) = - \frac{\varepsilon}{T}\partial_\tau g - \frac{\varepsilon R}{1-R} 
\left ( g + I f(gf - \partial_\sigma f ) \right )
\label{eq:gsigma_full}
\end{equation}

Once again we can remove the time derivative with the help of Eq.~(\ref{eq:gtau})
\begin{eqnarray}
g_\sigma &=& I(1-f^2) - \frac{\varepsilon}{T}\left [
I \left( f^2 -2 \widehat{g f^2}+ C_2 \right) \right ]   \nonumber \\
&& - \frac{\varepsilon R}{1-R} \left ( g + I f(gf - \partial_\sigma f ) \right )
\label{eq:finalg}
\end{eqnarray}

\section{Energy evolution}
\label{app:etot}
In order to derive the energy evolution equations it is convenient rewrite Eqs.~(\ref{eq:finalgs},\ref{eq:finalfs2}) 
by introducing $f = \ln s$,
\begin{equation}
s_\tau = g  -\Gamma (g_\sigma  +g s_\sigma ) + 
\Gamma (s_\sigma^2+s_{\sigma\sigma} )
\label{eq:ss}
\end{equation}
\begin{equation}
g_\sigma = I(1-{\rm e}^{2s}) - \Gamma \left ( g + I (g  - s_\sigma ){\rm e}^{2s} \right ) \; ,
\label{eq:gs}
\end{equation}

\subsection{Kinetic energy}

From the definition (\ref{eq:ekin}) of $E_K$,
\begin{equation}
I (E_K)_\tau = \langle s_\sigma s_{\tau\sigma} \rangle \, .
\end{equation}
From the $s$ evolution equation (\ref{eq:ss})
\begin{eqnarray}
s_{\tau\sigma} &=&  g_\sigma   
-\Gamma (g_{\sigma\sigma} + g_\sigma s_\sigma + g s_{\sigma\sigma}) \nonumber \\
&&  + \Gamma (2 s_\sigma s_{\sigma\sigma}+s_{\sigma\sigma\sigma})  \; .
\end{eqnarray}
Hence,
\begin{eqnarray}
I (E_K)_\tau  &=& 
 \langle g_\sigma  s_\sigma \rangle  - \nonumber \\
&& \Gamma ( \langle g_{\sigma\sigma}s_\sigma \rangle  + \langle g_\sigma s_\sigma^2\rangle +
\langle g s_{\sigma\sigma}s_\sigma \rangle) + \\
&&  \Gamma ( 2 \langle s_\sigma^2  s_{\sigma\sigma}\rangle +\langle s_{\sigma\sigma\sigma} s_\sigma\rangle )  \nonumber
\end{eqnarray}
The second last term vanishes
since it is the average of the derivative of a strictly periodic functions and we can write,
\begin{eqnarray}
I (E_K)_\tau &=& \langle g_\sigma s_\sigma \rangle - \label{eq:ekdot1} \\
&&\Gamma ( \langle g_{\sigma\sigma}s_\sigma \rangle + \langle g_\sigma s_\sigma^2 \rangle  +
\langle g s_{\sigma\sigma}s_\sigma \rangle - \langle s_{\sigma\sigma\sigma} s_\sigma \rangle )  \nonumber
\end{eqnarray}
Now, from Eq.~(\ref{eq:gs})  
\begin{eqnarray}
\langle g_\sigma s_\sigma\rangle &=& \langle I(1-{\rm e}^{2s})s_\sigma \rangle - 
\Gamma \langle ( g + I (g  - s_\sigma ){\rm e}^{2s} )s_\sigma \rangle = \nonumber \\
&&  - \Gamma \langle ( g + I (g  - s_\sigma ){\rm e}^{2s} )s_\sigma\rangle
\end{eqnarray}
By inserting this expression into Eq.~(\ref{eq:ekdot1})
\begin{eqnarray}
I (E_K)_\tau &=&
-\Gamma [ \langle g_{\sigma\sigma}s_\sigma \rangle + \langle g_\sigma s_\sigma^2\rangle + 
\langle g s_{\sigma\sigma}s_\sigma \rangle - \\
&& \langle s_{\sigma\sigma\sigma} s_\sigma \rangle +
(1+I)\langle g s_\sigma \rangle -I\langle {\rm e}^{2s}s_\sigma^2 \rangle] ~. \nonumber
\end{eqnarray}
This equation  confirms that the kinetic energy is indeed a constant of motion in the limit $\Gamma = 0$
and provides an expression for its derivative, which is explicitly proportional to $\Gamma$.
The equation can be simplified by inserting first-order approximations in the r.h.s.
For instance,
\begin{equation}
\langle g_{\sigma\sigma}s_\sigma \rangle = -2I\langle {\rm e}^{2s}s_\sigma^2 \rangle \, .
\end{equation}
and
\begin{equation}
\langle g_\sigma s_\sigma^2\rangle = I\langle s_\sigma^2\rangle-I\langle {\rm e}^{2s}s_\sigma^2 \rangle
\end{equation}
As a result,
\begin{eqnarray}
I (E_K)_\tau  &=& -\Gamma [I\langle s_\sigma^2\rangle 
-4I \langle {\rm e}^{2s}s_\sigma^2 \rangle + \langle g s_{\sigma\sigma}s_\sigma\rangle 
- \nonumber \\
&& \langle s_{\sigma\sigma\sigma} s_\sigma \rangle +(1+I) \langle g s_\sigma \rangle ] ~.
\end{eqnarray}
One can eliminate the variable $g$ by using the following two transformations
\begin{eqnarray*}
&& 2 \int g s_{\sigma\sigma}s_\sigma~d\sigma  = \int g ( s_\sigma^2)_\sigma ~d\sigma = 
 gs_\sigma^2-\int s_\sigma^2 g_\sigma~d\sigma  \nonumber \\
&& = gs_\sigma^2-I\int s_\sigma^2 (1-{\rm e}^{2s})~d\sigma ~,
\end{eqnarray*}
and 
\begin{equation*}
\int g s_\sigma~d\sigma = gs -\int s g_\sigma~d\sigma = gs-I\int s (1-{\rm e}^{2s})~d\sigma  ~,
\end{equation*}
The final result is
\begin{equation}
\frac{(E_K)_\tau}{\Gamma} =
-\frac{1}{2} \langle s_\sigma^2 \rangle  +\frac{7}{2}\langle {\rm e}^{2s}s_\sigma^2 \rangle -
(1+I)\langle s(e^{2s}-1) \rangle - 
\frac{1}{I} \langle s^2_{\sigma\sigma} \rangle ~.
\label{eq:ekindot_fin}
\end{equation}
where we have also made use of the identity
$\langle s_{\sigma\sigma\sigma} s_\sigma \rangle = -\langle s_{\sigma\sigma}^2\rangle$,
(proved, integrating by parts).

\subsection{Potential-energy evolution}

From the definition (\ref{eq:epon}) of the potential energy, it follows
\begin{equation}
(E_P)_\tau = 2\langle (\mathrm{e}^{2s}-1)s_\tau \rangle ~,
\end{equation}
and then
\begin{equation}
\frac{1}{2}(E_P)_\tau = 
\langle (\mathrm{e}^{2s}-1) g \rangle  
+ \Gamma \left [
-\langle \mathrm{e}^{2s}g_\sigma \rangle
- \langle \mathrm{e}^{2s}g s_\sigma \rangle 
+ \langle \mathrm{e}^{2s}s_\sigma^2 \rangle 
+ \langle \mathrm{e}^{2s}s_{\sigma\sigma} \rangle 
+ \langle g s_\sigma \rangle - \langle s_\sigma^2 \rangle \right ]  ~.
\end{equation}
At first order
\begin{equation}
I \langle (\mathrm{e}^{2s}-1) g \rangle =  
- \Gamma \left( \langle  g^2 \rangle + I \langle g^2 e^{2s}\rangle  
- I \langle g s_\sigma e^{2s}\rangle \right) ~. 
\end{equation}
Hence,
\begin{equation}
\frac{1}{2\Gamma}(E_P)_\tau = 
- \frac{1}{I} \langle  g^2 \rangle 
- \langle g^2 e^{2s}\rangle  
+\langle \mathrm{e}^{2s}(\mathrm{e}^{2s}-1) \rangle
-\langle \mathrm{e}^{2s}s_\sigma^2 \rangle
+ \langle g s_\sigma \rangle - \langle s_\sigma^2 \rangle   ~.
\label{eq:epotdot_fin}
\end{equation}

 
\bibliographystyle{apsrev4-2} 

\begin{thebibliography}{24}%
\makeatletter
\providecommand \@ifxundefined [1]{%
 \@ifx{#1\undefined}
}%
\providecommand \@ifnum [1]{%
 \ifnum #1\expandafter \@firstoftwo
 \else \expandafter \@secondoftwo
 \fi
}%
\providecommand \@ifx [1]{%
 \ifx #1\expandafter \@firstoftwo
 \else \expandafter \@secondoftwo
 \fi
}%
\providecommand \natexlab [1]{#1}%
\providecommand \enquote  [1]{``#1''}%
\providecommand \bibnamefont  [1]{#1}%
\providecommand \bibfnamefont [1]{#1}%
\providecommand \citenamefont [1]{#1}%
\providecommand \href@noop [0]{\@secondoftwo}%
\providecommand \href [0]{\begingroup \@sanitize@url \@href}%
\providecommand \@href[1]{\@@startlink{#1}\@@href}%
\providecommand \@@href[1]{\endgroup#1\@@endlink}%
\providecommand \@sanitize@url [0]{\catcode `\\12\catcode `\$12\catcode
  `\&12\catcode `\#12\catcode `\^12\catcode `\_12\catcode `\%12\relax}%
\providecommand \@@startlink[1]{}%
\providecommand \@@endlink[0]{}%
\providecommand \url  [0]{\begingroup\@sanitize@url \@url }%
\providecommand \@url [1]{\endgroup\@href {#1}{\urlprefix }}%
\providecommand \urlprefix  [0]{URL }%
\providecommand \Eprint [0]{\href }%
\providecommand \doibase [0]{https://doi.org/}%
\providecommand \selectlanguage [0]{\@gobble}%
\providecommand \bibinfo  [0]{\@secondoftwo}%
\providecommand \bibfield  [0]{\@secondoftwo}%
\providecommand \translation [1]{[#1]}%
\providecommand \BibitemOpen [0]{}%
\providecommand \bibitemStop [0]{}%
\providecommand \bibitemNoStop [0]{.\EOS\space}%
\providecommand \EOS [0]{\spacefactor3000\relax}%
\providecommand \BibitemShut  [1]{\csname bibitem#1\endcsname}%
\let\auto@bib@innerbib\@empty
\bibitem [{\citenamefont {Arecchi}\ and\ \citenamefont
  {Bonifacio}(1965)}]{Arecchi1965}%
  \BibitemOpen
  \bibfield  {author} {\bibinfo {author} {\bibfnamefont {F.}~\bibnamefont
  {Arecchi}}\ and\ \bibinfo {author} {\bibfnamefont {R.}~\bibnamefont
  {Bonifacio}},\ }\href {https://doi.org/10.1109/JQE.1965.1072212} {\bibfield
  {journal} {\bibinfo  {journal} {IEEE J. Quantum Electron.}\ }\textbf
  {\bibinfo {volume} {1}},\ \bibinfo {pages} {169} (\bibinfo {year}
  {1965})}\BibitemShut {NoStop}%
\bibitem [{\citenamefont {Risken}\ and\ \citenamefont
  {Nummedal}(1968{\natexlab{a}})}]{Risken1968}%
  \BibitemOpen
  \bibfield  {author} {\bibinfo {author} {\bibfnamefont {H.}~\bibnamefont
  {Risken}}\ and\ \bibinfo {author} {\bibfnamefont {K.}~\bibnamefont
  {Nummedal}},\ }\href {https://doi.org/10.1063/1.1655817} {\bibfield
  {journal} {\bibinfo  {journal} {J. Appl. Phys.}\ }\textbf {\bibinfo {volume}
  {39}},\ \bibinfo {pages} {4662} (\bibinfo {year}
  {1968}{\natexlab{a}})}\BibitemShut {NoStop}%
\bibitem [{\citenamefont {Risken}\ and\ \citenamefont
  {Nummedal}(1968{\natexlab{b}})}]{Risken1968a}%
  \BibitemOpen
  \bibfield  {author} {\bibinfo {author} {\bibfnamefont {H.}~\bibnamefont
  {Risken}}\ and\ \bibinfo {author} {\bibfnamefont {K.}~\bibnamefont
  {Nummedal}},\ }\href {https://doi.org/10.1016/0375-9601(68)90646-4}
  {\bibfield  {journal} {\bibinfo  {journal} {Phys. Lett. A}\ }\textbf
  {\bibinfo {volume} {26}},\ \bibinfo {pages} {275} (\bibinfo {year}
  {1968}{\natexlab{b}})}\BibitemShut {NoStop}%
\bibitem [{\citenamefont {Graham}\ and\ \citenamefont
  {Haken}(1968)}]{Graham1968}%
  \BibitemOpen
  \bibfield  {author} {\bibinfo {author} {\bibfnamefont {R.}~\bibnamefont
  {Graham}}\ and\ \bibinfo {author} {\bibfnamefont {H.}~\bibnamefont {Haken}},\
  }\href {https://doi.org/10.1007/BF01405384} {\bibfield  {journal} {\bibinfo
  {journal} {Zeitschrift f{\"{u}}r Phys. A Hadron. Nucl.}\ }\textbf {\bibinfo
  {volume} {213}},\ \bibinfo {pages} {420} (\bibinfo {year}
  {1968})}\BibitemShut {NoStop}%
\bibitem [{\citenamefont {Milonni}\ \emph {et~al.}(1987)\citenamefont
  {Milonni}, \citenamefont {Shih},\ and\ \citenamefont
  {Ackerhalt}}]{Milonni1987}%
  \BibitemOpen
  \bibfield  {author} {\bibinfo {author} {\bibfnamefont {P.~W.}\ \bibnamefont
  {Milonni}}, \bibinfo {author} {\bibfnamefont {M.-L.}\ \bibnamefont {Shih}},\
  and\ \bibinfo {author} {\bibfnamefont {J.~R.}\ \bibnamefont {Ackerhalt}},\
  }\href {https://doi.org/10.1142/0323} {\emph {\bibinfo {title} {{Chaos in
  Laser-Matter Interactions}}}},\ \bibinfo {series} {World Scientific Lecture
  Notes in Physics}, Vol.~\bibinfo {volume} {6}\ (\bibinfo  {publisher} {World
  Scientific},\ \bibinfo {year} {1987})\BibitemShut {NoStop}%
\bibitem [{\citenamefont {Casini}\ \emph {et~al.}(1997)\citenamefont {Casini},
  \citenamefont {D'Alessandro},\ and\ \citenamefont {Politi}}]{Casini1997}%
  \BibitemOpen
  \bibfield  {author} {\bibinfo {author} {\bibfnamefont {D.}~\bibnamefont
  {Casini}}, \bibinfo {author} {\bibfnamefont {G.}~\bibnamefont
  {D'Alessandro}},\ and\ \bibinfo {author} {\bibfnamefont {A.}~\bibnamefont
  {Politi}},\ }\href {https://doi.org/10.1103/PhysRevA.55.751} {\bibfield
  {journal} {\bibinfo  {journal} {Phys. Rev. A}\ }\textbf {\bibinfo {volume}
  {55}},\ \bibinfo {pages} {751} (\bibinfo {year} {1997})}\BibitemShut
  {NoStop}%
\bibitem [{\citenamefont {Agrawal}\ and\ \citenamefont
  {Dutta}(2013)}]{agrawal2013}%
  \BibitemOpen
  \bibfield  {author} {\bibinfo {author} {\bibfnamefont {G.}~\bibnamefont
  {Agrawal}}\ and\ \bibinfo {author} {\bibfnamefont {N.}~\bibnamefont
  {Dutta}},\ }\href {https://books.google.it/books?id=bInTBwAAQBAJ} {\emph
  {\bibinfo {title} {Semiconductor Lasers}}}\ (\bibinfo  {publisher} {Springer
  US},\ \bibinfo {year} {2013})\BibitemShut {NoStop}%
\bibitem [{\citenamefont {Turitsyna}\ \emph {et~al.}(2013)\citenamefont
  {Turitsyna}, \citenamefont {Smirnov}, \citenamefont {Sugavanam},
  \citenamefont {Tarasov}, \citenamefont {Shu}, \citenamefont {Babin},
  \citenamefont {{E. V.}}, \citenamefont {Churkin}, \citenamefont {Falkovich},\
  and\ \citenamefont {Turitsyn}}]{Turitsyna2013}%
  \BibitemOpen
  \bibfield  {author} {\bibinfo {author} {\bibfnamefont {E.~G.}\ \bibnamefont
  {Turitsyna}}, \bibinfo {author} {\bibfnamefont {S.~V.}\ \bibnamefont
  {Smirnov}}, \bibinfo {author} {\bibfnamefont {S.}~\bibnamefont {Sugavanam}},
  \bibinfo {author} {\bibfnamefont {N.}~\bibnamefont {Tarasov}}, \bibinfo
  {author} {\bibfnamefont {X.}~\bibnamefont {Shu}}, \bibinfo {author}
  {\bibfnamefont {S.~A.}\ \bibnamefont {Babin}}, \bibinfo {author}
  {\bibfnamefont {P.}~\bibnamefont {{E. V.}}}, \bibinfo {author} {\bibfnamefont
  {D.~V.}\ \bibnamefont {Churkin}}, \bibinfo {author} {\bibfnamefont
  {G.}~\bibnamefont {Falkovich}},\ and\ \bibinfo {author} {\bibfnamefont
  {S.~K.}\ \bibnamefont {Turitsyn}},\ }\href
  {http://dx.doi.org/10.1038/nphoton.2013.246} {\bibfield  {journal} {\bibinfo
  {journal} {Nat. Photon.}\ }\textbf {\bibinfo {volume} {7}},\ \bibinfo {pages}
  {783} (\bibinfo {year} {2013})}\BibitemShut {NoStop}%
\bibitem [{\citenamefont {Rogers}\ \emph {et~al.}(2005)\citenamefont {Rogers},
  \citenamefont {Peles},\ and\ \citenamefont {Wiesenfeld}}]{Rogers2005}%
  \BibitemOpen
  \bibfield  {author} {\bibinfo {author} {\bibfnamefont {J.}~\bibnamefont
  {Rogers}}, \bibinfo {author} {\bibfnamefont {S.}~\bibnamefont {Peles}},\ and\
  \bibinfo {author} {\bibfnamefont {K.}~\bibnamefont {Wiesenfeld}},\ }\href
  {https://doi.org/10.1109/JQE.2005.847545} {\bibfield  {journal} {\bibinfo
  {journal} {IEEE Journal of Quantum Electronics}\ }\textbf {\bibinfo {volume}
  {41}},\ \bibinfo {pages} {767} (\bibinfo {year} {2005})}\BibitemShut
  {NoStop}%
\bibitem [{\citenamefont {Abarbanel}\ \emph {et~al.}(1999)\citenamefont
  {Abarbanel}, \citenamefont {Kennel}, \citenamefont {Buhl},\ and\
  \citenamefont {Lewis}}]{Abarbanel1999}%
  \BibitemOpen
  \bibfield  {author} {\bibinfo {author} {\bibfnamefont {H.~D.}\ \bibnamefont
  {Abarbanel}}, \bibinfo {author} {\bibfnamefont {M.~B.}\ \bibnamefont
  {Kennel}}, \bibinfo {author} {\bibfnamefont {M.}~\bibnamefont {Buhl}},\ and\
  \bibinfo {author} {\bibfnamefont {C.~T.}\ \bibnamefont {Lewis}},\ }\href
  {https://doi.org/10.1103/PhysRevA.60.2360} {\bibfield  {journal} {\bibinfo
  {journal} {Physical Review A}\ }\textbf {\bibinfo {volume} {60}},\ \bibinfo
  {pages} {2360} (\bibinfo {year} {1999})}\BibitemShut {NoStop}%
\bibitem [{\citenamefont {Ray}\ \emph {et~al.}(2008)\citenamefont {Ray},
  \citenamefont {Wiesenfeld},\ and\ \citenamefont {Rogers}}]{Ray2008}%
  \BibitemOpen
  \bibfield  {author} {\bibinfo {author} {\bibfnamefont {W.}~\bibnamefont
  {Ray}}, \bibinfo {author} {\bibfnamefont {K.}~\bibnamefont {Wiesenfeld}},\
  and\ \bibinfo {author} {\bibfnamefont {J.~L.}\ \bibnamefont {Rogers}},\
  }\href {https://doi.org/10.1103/PhysRevE.78.046203} {\bibfield  {journal}
  {\bibinfo  {journal} {Physical Review E}\ }\textbf {\bibinfo {volume} {78}},\
  \bibinfo {pages} {1} (\bibinfo {year} {2008})}\BibitemShut {NoStop}%
\bibitem [{\citenamefont {Haus}(1975)}]{Haus1975}%
  \BibitemOpen
  \bibfield  {author} {\bibinfo {author} {\bibfnamefont {H.}~\bibnamefont
  {Haus}},\ }\href {https://doi.org/10.1109/JQE.1975.1068636} {\bibfield
  {journal} {\bibinfo  {journal} {IEEE Journal of Quantum Electronics}\
  }\textbf {\bibinfo {volume} {11}},\ \bibinfo {pages} {323} (\bibinfo {year}
  {1975})}\BibitemShut {NoStop}%
\bibitem [{\citenamefont {Perego}\ \emph {et~al.}(2020)\citenamefont {Perego},
  \citenamefont {Garbin}, \citenamefont {Gustave}, \citenamefont {Barland},
  \citenamefont {Prati},\ and\ \citenamefont
  {de~Valc{\'{a}}rcel}}]{Perego2020}%
  \BibitemOpen
  \bibfield  {author} {\bibinfo {author} {\bibfnamefont {A.~M.}\ \bibnamefont
  {Perego}}, \bibinfo {author} {\bibfnamefont {B.}~\bibnamefont {Garbin}},
  \bibinfo {author} {\bibfnamefont {F.}~\bibnamefont {Gustave}}, \bibinfo
  {author} {\bibfnamefont {S.}~\bibnamefont {Barland}}, \bibinfo {author}
  {\bibfnamefont {F.}~\bibnamefont {Prati}},\ and\ \bibinfo {author}
  {\bibfnamefont {G.~J.}\ \bibnamefont {de~Valc{\'{a}}rcel}},\ }\href
  {https://doi.org/10.1038/s41467-019-14013-4} {\bibfield  {journal} {\bibinfo
  {journal} {Nat. Commun.}\ }\textbf {\bibinfo {volume} {11}},\ \bibinfo
  {pages} {311} (\bibinfo {year} {2020})}\BibitemShut {NoStop}%
\bibitem [{\citenamefont {Vladimirov}\ and\ \citenamefont
  {Turaev}(2005)}]{Vladimirov2005}%
  \BibitemOpen
  \bibfield  {author} {\bibinfo {author} {\bibfnamefont {A.~G.}\ \bibnamefont
  {Vladimirov}}\ and\ \bibinfo {author} {\bibfnamefont {D.}~\bibnamefont
  {Turaev}},\ }\href {https://doi.org/10.1103/PhysRevA.72.033808} {\bibfield
  {journal} {\bibinfo  {journal} {Physical Review A}\ }\textbf {\bibinfo
  {volume} {72}},\ \bibinfo {pages} {033808} (\bibinfo {year}
  {2005})}\BibitemShut {NoStop}%
\bibitem [{\citenamefont {Giacomelli}\ \emph {et~al.}(2021)\citenamefont
  {Giacomelli}, \citenamefont {Yanchuk},\ and\ \citenamefont
  {Politi}}]{Giacomelli2021}%
  \BibitemOpen
  \bibfield  {author} {\bibinfo {author} {\bibfnamefont {G.}~\bibnamefont
  {Giacomelli}}, \bibinfo {author} {\bibfnamefont {S.}~\bibnamefont
  {Yanchuk}},\ and\ \bibinfo {author} {\bibfnamefont {A.}~\bibnamefont
  {Politi}},\ }\href {https://doi.org/10.1103/PhysRevA.104.053521} {\bibfield
  {journal} {\bibinfo  {journal} {Phys. Rev. A}\ }\textbf {\bibinfo {volume}
  {104}},\ \bibinfo {pages} {053521} (\bibinfo {year} {2021})}\BibitemShut
  {NoStop}%
\bibitem [{\citenamefont {Politi}\ \emph {et~al.}(2023)\citenamefont {Politi},
  \citenamefont {Yanchuk},\ and\ \citenamefont {Giacomelli}}]{Politi2023}%
  \BibitemOpen
  \bibfield  {author} {\bibinfo {author} {\bibfnamefont {A.}~\bibnamefont
  {Politi}}, \bibinfo {author} {\bibfnamefont {S.}~\bibnamefont {Yanchuk}},\
  and\ \bibinfo {author} {\bibfnamefont {G.}~\bibnamefont {Giacomelli}},\
  }\href@noop {} {\bibfield  {journal} {\bibinfo  {journal} {Physical Review
  Research}\ }\textbf {\bibinfo {volume} {5}},\ \bibinfo {pages} {023059}
  (\bibinfo {year} {2023})}\BibitemShut {NoStop}%
\bibitem [{\citenamefont {Toda}(1975)}]{Toda1975}%
  \BibitemOpen
  \bibfield  {author} {\bibinfo {author} {\bibfnamefont {M.}~\bibnamefont
  {Toda}},\ }\href
  {https://doi.org/https://doi.org/10.1016/0370-1573(75)90018-6} {\bibfield
  {journal} {\bibinfo  {journal} {Physics Reports}\ }\textbf {\bibinfo {volume}
  {18}},\ \bibinfo {pages} {1} (\bibinfo {year} {1975})}\BibitemShut {NoStop}%
\bibitem [{\citenamefont {Bainov}\ and\ \citenamefont
  {Mishev}(1991)}]{bainov1991oscillation}%
  \BibitemOpen
  \bibfield  {author} {\bibinfo {author} {\bibfnamefont {D.}~\bibnamefont
  {Bainov}}\ and\ \bibinfo {author} {\bibfnamefont {D.}~\bibnamefont
  {Mishev}},\ }\href {https://books.google.it/books?id=VMKxQlPvjjEC} {\emph
  {\bibinfo {title} {Oscillation Theory for Neutral Differential Equations with
  Delay}}}\ (\bibinfo  {publisher} {Taylor \& Francis},\ \bibinfo {year}
  {1991})\BibitemShut {NoStop}%
\bibitem [{\citenamefont {Arecchi}\ \emph {et~al.}(1992)\citenamefont
  {Arecchi}, \citenamefont {Giacomelli}, \citenamefont {Lapucci},\ and\
  \citenamefont {Meucci}}]{Arecchi1992}%
  \BibitemOpen
  \bibfield  {author} {\bibinfo {author} {\bibfnamefont {F.~T.}\ \bibnamefont
  {Arecchi}}, \bibinfo {author} {\bibfnamefont {G.}~\bibnamefont {Giacomelli}},
  \bibinfo {author} {\bibfnamefont {A.}~\bibnamefont {Lapucci}},\ and\ \bibinfo
  {author} {\bibfnamefont {R.}~\bibnamefont {Meucci}},\ }\href
  {https://doi.org/10.1103/PhysRevA.45.R4225} {\bibfield  {journal} {\bibinfo
  {journal} {Phys. Rev. A}\ }\textbf {\bibinfo {volume} {45}},\ \bibinfo
  {pages} {R4225} (\bibinfo {year} {1992})}\BibitemShut {NoStop}%
\bibitem [{\citenamefont {Giacomelli}\ and\ \citenamefont
  {Politi}(1996)}]{Giacomelli1996}%
  \BibitemOpen
  \bibfield  {author} {\bibinfo {author} {\bibfnamefont {G.}~\bibnamefont
  {Giacomelli}}\ and\ \bibinfo {author} {\bibfnamefont {A.}~\bibnamefont
  {Politi}},\ }\href {https://doi.org/10.1103/PhysRevLett.76.2686} {\bibfield
  {journal} {\bibinfo  {journal} {Physical Review Letters}\ }\textbf {\bibinfo
  {volume} {76}},\ \bibinfo {pages} {2686} (\bibinfo {year}
  {1996})}\BibitemShut {NoStop}%
\bibitem [{\citenamefont {Lugiato}\ \emph {et~al.}(1986)\citenamefont
  {Lugiato}, \citenamefont {Narducci},\ and\ \citenamefont
  {Squicciarini}}]{Lugiato1986}%
  \BibitemOpen
  \bibfield  {author} {\bibinfo {author} {\bibfnamefont {L.~A.}\ \bibnamefont
  {Lugiato}}, \bibinfo {author} {\bibfnamefont {L.~M.}\ \bibnamefont
  {Narducci}},\ and\ \bibinfo {author} {\bibfnamefont {M.~F.}\ \bibnamefont
  {Squicciarini}},\ }\href {https://doi.org/10.1103/PhysRevA.34.3101}
  {\bibfield  {journal} {\bibinfo  {journal} {Phys. Rev. A}\ }\textbf {\bibinfo
  {volume} {34}},\ \bibinfo {pages} {3101} (\bibinfo {year}
  {1986})}\BibitemShut {NoStop}%
\bibitem [{Note1()}]{Note1}%
  \BibitemOpen
  \bibinfo {note} {Unimodal and bimodal refer to the $s = \protect \qopname
  \relax o{ln}f$ representation}\BibitemShut {NoStop}%
\bibitem [{\citenamefont {Oppo}\ and\ \citenamefont {Politi}(1986)}]{Oppo1986}%
  \BibitemOpen
  \bibfield  {author} {\bibinfo {author} {\bibfnamefont {G.~L.}\ \bibnamefont
  {Oppo}}\ and\ \bibinfo {author} {\bibfnamefont {A.}~\bibnamefont {Politi}},\
  }\href {https://doi.org/10.1209/0295-5075/1/11/002} {\bibfield  {journal}
  {\bibinfo  {journal} {Europhysics Letters ({EPL})}\ }\textbf {\bibinfo
  {volume} {1}},\ \bibinfo {pages} {549} (\bibinfo {year} {1986})}\BibitemShut
  {NoStop}%
\bibitem [{\citenamefont {Oppo}\ and\ \citenamefont {Politi}(1989)}]{Oppo89}%
  \BibitemOpen
  \bibfield  {author} {\bibinfo {author} {\bibfnamefont {G.-L.}\ \bibnamefont
  {Oppo}}\ and\ \bibinfo {author} {\bibfnamefont {A.}~\bibnamefont {Politi}},\
  }\href {https://doi.org/10.1103/PhysRevA.40.1422} {\bibfield  {journal}
  {\bibinfo  {journal} {Phys. Rev. A}\ }\textbf {\bibinfo {volume} {40}},\
  \bibinfo {pages} {1422} (\bibinfo {year} {1989})}\BibitemShut {NoStop}%

\end{thebibliography}

%

\end{document}